\documentclass[aps, prr, twocolumn, notitlepage,superscriptaddress,nofootinbib,floatfix   
]{revtex4-2}
\usepackage{graphicx}
\usepackage{bm}
\usepackage{amsmath}
\usepackage{amssymb}
\usepackage[caption=false,lofdepth,lotdepth]{subfig}
\usepackage{xcolor}
\usepackage[normalem]{ulem}
\usepackage{cancel}
\usepackage{hyperref}   
\hypersetup{%
	pdfpagemode=FullScreen,
	pdfstartpage=1,
	pdfmenubar=true,
	pdftoolbar=true,
	colorlinks = true,
	linkcolor=blue,
	citecolor=blue,
	urlcolor=blue,
	bookmarksopen=false
}
\usepackage{float}
\allowdisplaybreaks

\begin{document}

\title{Massive-vortex realization of a Bosonic Josephson Junction}
\author{Alice Bellettini}
\email{alice.bellettini@polito.it}
\affiliation{Department of Applied Science and Technology, Politecnico di Torino, 10129 Torino, Italy}

\author{Andrea Richaud}
\affiliation{Departament de F\'isica, Universitat Polit\`ecnica de Catalunya, Campus Nord B4-B5, E-08034 Barcelona, Spain}

\author{Vittorio Penna}
\affiliation{Department of Applied Science and Technology, Politecnico di Torino, 10129 Torino, Italy}

\date{\today}

\begin{abstract}
    We study the mass exchange between two rotating, quantum massive vortices in a two-component Bose-Einstein condensate.
    The vortices, in the majority component, exhibit a filled core, where the in-filling minority component undergoes a quantum tunneling effect.
    Remarkably, we observe that the two-vortex system features stable Josephson oscillations, as well as all the nonlinear phenomena, including the macroscopic quantum self-trapping, that characterize a Bosonic Josephson Junction (BJJ).
    We propose an analytical model for describing the inter-vortex tunneling, obtained by implementing the a coherent-state representation of the two-mode Bose-Hubbard model. This allows us to give the explicit expression of the model's parameters in terms of the physical macroscopic parameters of the two-vortex system.
    The comparison of the dynamical scenario predicted by the model with that emerging from the Gross-Pitaevskii equations
    is very good for sufficiently small particle numbers, 
    while at larger particle numbers it grows less precise, presumably due to the partial exclusion of the many-body interactions from our model.
    The definition of an effective self-interaction parameter allows us to include the many-body effects, thus restoring
    a quite good agreement with the numerical results. 
    Interestingly, the
    recognition of the bosonic Josephson dynamics
    paves the way to the investigation of new dynamical behaviors in multi-vortex configurations.    
\end{abstract}

\maketitle

\section{Introduction}

The recent research in quantum massive vortices \cite{Matthews1999, Anderson2000} has unveiled novel and interesting phenomena, such as the formation of stable giant vortices \cite{Kuopanportti2015, Patrick2023, Richaud2023} or exotic vortex lattices \cite{Mueller2002, Schweikhard2004, Kuopanportti2012}, vortex collisions \cite{Richaud2023},
vortex splitting \cite{Kuopanportti2019}, and a richer dynamics with respect to massless vortices which includes rotational states of asymmetric vortex pairs and multi-mode oscillatory behaviors supported by inertial effects \cite{Richaud2020, Bellettini2023, Bellettini2024}. The larger manipulability of quantum gases compared to liquid helium has also opened the path to superfluidity experiments in different trap geometries \cite{White2006, Navon2021, Caldara2023} and even topologies \cite{Lundblad2019, Guo2022, Wolf2022, Móller2020, Caracanhas2022}. 
Dilute, ultracold quantum gases are in fact confined via optical potentials, where the pinning of a vortex to a desired position is also possible\cite{Davis2009, Samson2016}. Such a framework enables experimental realizations of different massive vortices configurations. 

Massive vortices can occur in the immiscibility regime of mixtures formed by two condensates
\cite{Gallemí2018,Kuopanportti2019}.
The dynamics of a binary mixture, well described by two coupled Gross-Pitaevskii equations (GPEs), within a mean-field picture, shows the formation of vortices in the majority component denoted by $a$.
These represent phase singularities characterized by the formation of a local density well at their cores \cite{Onsager1949, Fiszdon1991}. 
The introduction of a second, minority component (denoted by $b$), such as a second atomic species or the same species associated with a different hyperfine state, shows that its accumulation at the vortex cores is favoured if the two components are immiscible \cite{Gallemí2018}, 
\cite{ McGee2001, Law2010}.

The resulting density peaks in the component $b$, playing the role of effective inertial masses,
allowed the derivation of a point-like picture of  massive vortices
\cite{Richaud2020,  Richaud2021}. In this framework, the vortex dynamics is described by Lorenz-like equations, and involves the motion of the vortices while excluding the time dependence of the masses occupying the vortex cores. In this paper we overcome this constraint and investigate vortex configurations where a mass exchange between the vortices can take place, caused by a tunneling effect.

The well-known Bose-Hubbard (BH) model describes the interwell tunneling effect of bosons that are trapped in arrays of potential wells. For this reason it provides the natural framework where the boson exchange between different vortex cores in a binary mixture can be realistically modelled. 

The simplest possible model, the two-well BH model, has received outstanding attention in the last three decades both due to its integrable character and because its nonlinear dynamics, in addition to many interesting properties, exhibits a profound similarity with the Josephson-junction phenomenology.
This aspect was emphasized by Milburn \textit{et al.} \cite{Milburn1997} who conducted a study on the mean-field dynamics of neutral atoms of a Bose-Einstein condensate (BEC) in a double-well potential highlighting the presence of regimes characterized by the self-trapping transition and the Josephson effect. Further studies of such system led to the definition of Bosonic Josephson Junction (BJJ) \cite{Smerzi1997, Raghavan1999}, and to the observation of the ac and dc Josephson effects \cite{Giovanazzi2000}.

The theoretical literature on BJJs is vast. Among many aspects and effects, it ranges from the coherent-state picture of BJJ dynamics \cite{FRANZOSI2000},
the study of the double-well dynamics with the GPE \cite{ana2006}, the number squeezing effect \cite{ming2008} and
exact multiconfigurational dynamical approaches \cite{alon2009},
to phase diffusion processes \cite{Chuchem2010}, the transition to dissipative and self-trapping regimes
\cite{Xhani2020}
and the realization of atomtronics devices \cite{Amico2017}.
The effect of a second component on the
Josephson oscillations \cite{Mazzarella2009} and on the spectral properties \cite{Lingua2016}
were also explored, for a binary condensate in a double-well potential, or in more complex geometries
\cite{Penna2017,Richaud2019,Arwas2015}.

On the experimental front, Ref. \cite{Cataliotti2001} reported the direct observation of oscillatory atomic currents in a chain of BJJs. However, the first realization of a single BJJ, with $^{87}\mathrm{Rb}$, was done by Albiez \textit{et al.} \cite{Albiez2005} a few years later (a review on BJJ experiments is carried out in Ref. \cite{Gati2007}). This validated the presence of tunneling oscillations and of the macroscopic quantum self-trapping, which was also observed in a one-dimensional periodic potential \cite{Anker2005}.
Afterwards, the creation of a BJJ with a binary condensate \cite{Spagnolli2017} allowed to highlight the dynamical effects due to the intraspecies interactions. 

In this paper, we show 
that it is possible to realize a double-well BH model, namely a BJJ, by means of a pair of two-dimensional (2D) quantum massive vortices occurring in the component $a$, when the tunneling process of $b$ between the vortex cores is taken into account. Note that the component $a$, hosting the vortices, acts as an effective potential for the component $b$.
Our results are supported by the numerical simulation of the coupled GPEs for the mixture.
Surprisingly, we find that the inter-vortex mass imbalance and phase-shift between the two peaks, as obtained from the
GPEs dynamics,
reproduce the BH phase portraits characterizing a double-well system, where the trajectories are both stable over time and robust against eventual $b$ leaks outside of the vortex cores.

To validate these findings we implement the space-mode approximation on the field Hamiltonian of the component $b$ and derive analytically
the two-mode BH model and the relevant mean-field version
within the coherent-state picture. This allows us to find the phase-space  portrait of the BJJ, well known in the literature
(see e.g. \cite{Raghavan1999},
\cite{FRANZOSI2000}). The comparison of the mean-field model with the phase portrait provided by the GPE simulations shows a remarkably good 
agreement. Interestingly, this result implies that vortices can be used, in place of optical potentials, to support the formation of a BJJ.
In passing, we note how the derivation of the two-mode BH Hamiltonian allows us to find analytical expressions for the interaction and tunneling \textit{parameters} directly related to the macroscopic physical parameters of the vortex pair.

Pola \textit{et al.} \cite{Pola2012} already 
investigated the possibility of creating a BJJ via a vortex \textit{dipole} hosting a solitonic component.
In their work, however, they observed the BJJ dynamics in the limit of a frozen vortex dipole and of a very few particles of the in-filling component. Conversely, in our paper, we prove the existence of a robust BJJ dynamics for long times and for a wide range of $b$-component particle number. 
In fact, the support of the BJJ is in our case a orbiting vortex pair, where the two vortex wells have the same circulation. 
Unlike in Ref. \cite{Pola2012}, our system, involving also large numbers of particles in the vortex-hosting component, does not present a significant breathing of the vortices.

This paper is articulated as follows.
In Section \ref{sec:GPEs} we motivate our work and introduce the GPEs describing the mixture and the vortex pair dynamics.
Following up on that, we present in Section \ref{sec:numerical_results} the numeric results
revealing the presence of a Bosonic Josephson Junction in the system of the massive vortex pair, proving its stability and robustness.
In Section \ref{sec:theoretical_modelling} we subsequently introduce the analytical framework connected to a Bosonic Josephson Junction, starting from a two-mode BH Hamiltonian and deriving then its mean-field version via a coherent-state variational approach. Here we also derive some explicit analytical expressions for the parameters of the Bose-Hubbard model in terms of the macroscopic parameters of the vortex-pair system. We then discuss the comparison of the numerics with the analytical model in Section \ref{sec:comparison}, and sum up the conclusions and outlooks in Section \ref{sec:conclusion}. Hereafter, for the sake of simplicity, we shall refer to the doubel-well or two-modes BH model with the term {\it BH dimer}.

\section{Gross-Pitaevskii dynamics and BJJ realization}
\label{sec:GPEs}

The current paper is prompted by Ref. \cite{Bellettini2024}
where we proved the existence of
asymmetric rotational states of a vortex pair involving two mass-imbalanced vortices with same circulation. Vortices rotate specularly around the origin at different radial positions. Our analytical solution relied on a point-like model, which is based on the time-dependent variational Lagrangian approach \cite{Kim2004, Richaud2020}.
We characterized all the possible configurations of the vortex pair, obtaining a rich diagram of the rotational states, also involving massless vortices.
Moreover, we discussed the dynamical stability of the solutions in presence of small perturbations. Here arose the intuition that an adiabatic variation of the vortex masses could be a phenomenon present in GPEs dynamics of the mixture. 
In this case, by switching to an suitable rotating frame of reference, a BJJ dynamics for the $b$-component could be retrieved.

We start here with the symmetric vortex pair, where the two vortices are identical and specularly rotating at the same distance $r_1$ from the origin. 
The properties of the condesates' mixture are well captured by two coupled Gross-Pitaevskii equations \cite{Pitaevskii2016}, describing the two ultracold dilute quantum gases

\begin{equation}
\begin{cases}
   i\hbar \dot{\psi}_a& =\bigg(\frac{g_a}{L_z}|\psi_a|^2+\frac{g_{ab}}{L_z}|\psi_b|^2 
 -\frac{\hbar^2 \nabla^2}{2 m_a}+V_{ext}  \bigg)\psi_a\\
 &\\
 i\hbar \dot{\psi}_b& =\bigg(\frac{g_b}{L_z}|\psi_b|^2+\frac{g_{ab}}{L_z}|\psi_a|^2 
 -\frac{\hbar^2 \nabla^2}{2 m_b}+V_{ext}   \bigg)\psi_b,
\end{cases}
\end{equation}

where $V_{ext} (\bm{r})=V_{ext}(\bm{r})$ is the external potential.
The order parameters $\psi_a=\psi_a(x,y,t)$ and $\psi_b=\psi_b(x,y,t)$ are such that 
$$
\int d^2 r\;|\psi_a|^2=N_a, \quad 
\int d^2r
\;|\psi_b|^2=N_b,$$ 
where $N_a$ and $N_b$ are the number of atoms of component $a$ and $b$, respectively, and represent two constant of motions.
For the numerical simulation of the GPEs we employ a
mixture of $^{23}\mathrm{Na}$ ($a$) and $^{39}\mathrm{K}$ ($b$) confined in a 2D disk of radius $R=50\,\mu m$.
Hence, $V_{ext}(\bm{r})$ is a rigid wall potential in correspondence of the disk boundary and zero inside the disc. Masses
$m_a$ and $m_b$ are respectively the atomic masses of $^{23}\mathrm{Na}$ and of $^{39}\mathrm{K}$.
The repulsive boson-boson interactions are
defined by $$
g_a=\frac{4\pi\hbar^2 a_a}{m_a}, 
\quad
g_b=\frac{4\pi\hbar^2 a_b}{m_b}, 
\quad
g_{ab}=\frac{2\pi\hbar^2 a_{ab}}{m_r},
$$
where $a_a\simeq 52.0\, a_0$, $a_b\simeq 7.6\, a_0$, $a_{ab}\simeq 24.3   \,a_0$ 
and $a_0$ is the Bohr radius. $a_a$ and $a_b$ are 
the intraspecies s-wave scattering lengths, while $a_{ab}$ is the scattering length between an atom of $a$ and an atom of $b$. Note that in our case $g_{ab}/\sqrt{g_a g_b}=1.26$, i.e. the two components are immiscible \cite{Gutierrez2021}. At higher values of this ratio, the density profile at the vortex sites gets further from a local harmonic potential, whereas in the miscible regime the two peaks grow delocalized from the vortex centers. The reduced mass 
$m_r$ is such that $1/m_r=1/m_a+1/m_b$. The effective thickness of the layer in the direction $z$ where the dynamics is frozen is 
$L_z = 2\times10^{-6}\;m$. The dimensional reduction of the GPEs leads to the effective 2D normalization of the parameters $g_i$ by $L_z$ and to planar densities.
In the following, $N_a$ and $N_b$ are the (conserved) total number of particles of the component $a$, respectively $b$ and we always take $N_a=10^5$ for the numerical simulations.

\section{Results}
\label{sec:numerical_results}

\begin{figure}
    \centering
    \includegraphics[width=0.4\textwidth]{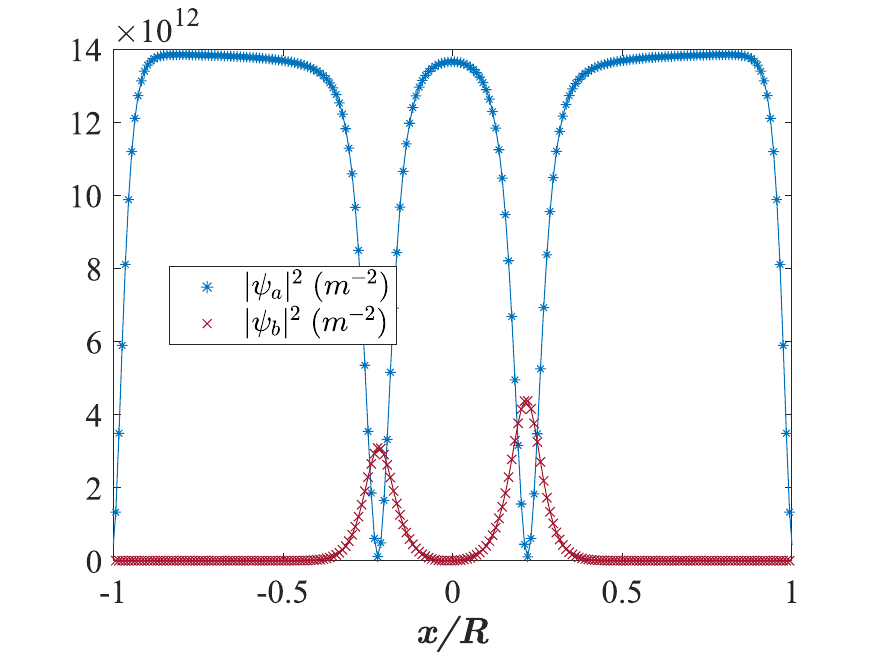}
    \caption{Density profiles $|\psi_a|^2$ and $|\psi_b|^2$ at $t=0$ along the $x$-direction and at $y=0$. The $a$ component hosts a vortex pair while the $b$ is the tunneling component, featuring a BJJ dynamics.}
    \label{fig:density_slice}
\end{figure}

The numerical simulations of the two co-rotating massive vortices show, with unexpected clarity, that the inter-vortex mass exchange essentially reproduces the BJJ
mechanism
(see Appendix \ref{app:code_description} for the code's description). 
The two vortices persistently orbit around the trap center keeping a constant distance (up to a small ripple of the order of $10^{-3}\;R$ that we neglect), and the local fraction of component $b$ is well trapped although
the $a$-density profile is only locally harmonic. This is visible in
Fig. \ref{fig:density_slice}, showing the density profiles of both the components $a$ and $b$.

\begin{figure}
    \centering
    \includegraphics[width=0.5\textwidth]{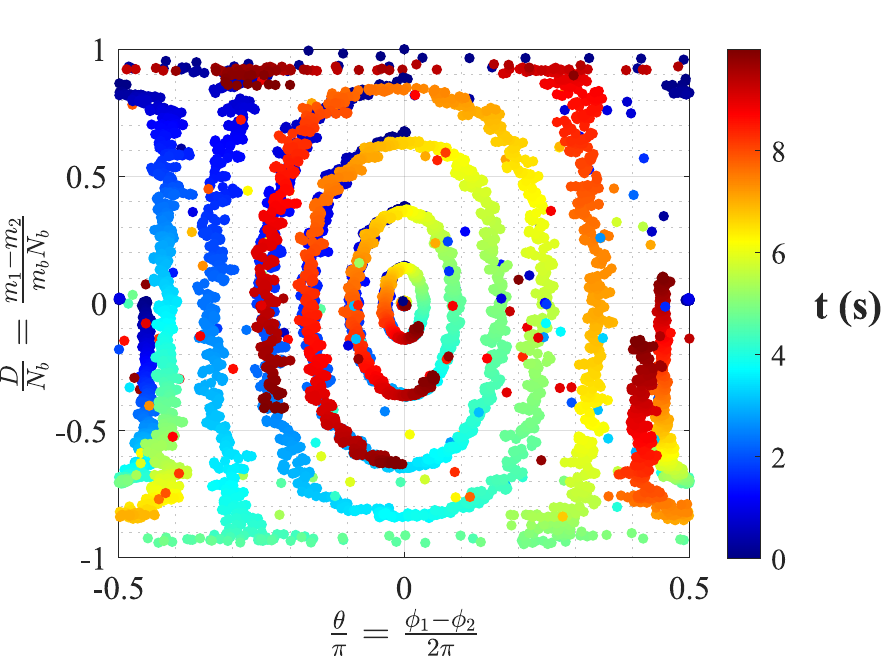}
    \caption{ GPE $(\theta, D)$ curves for a system with $N_b=100$ and $r_1/R\simeq 0.22$. The collection of the extracted trajectories resembles the BH-dimer dynamics in a specific regime. The curves are parameterized by the time. The two fixed points of the BJJ model are reproduced along the axis $D=0$.}
    \label{fig:pp_d}
\end{figure}

\begin{figure}
    \centering
    \includegraphics[width=0.5\textwidth]{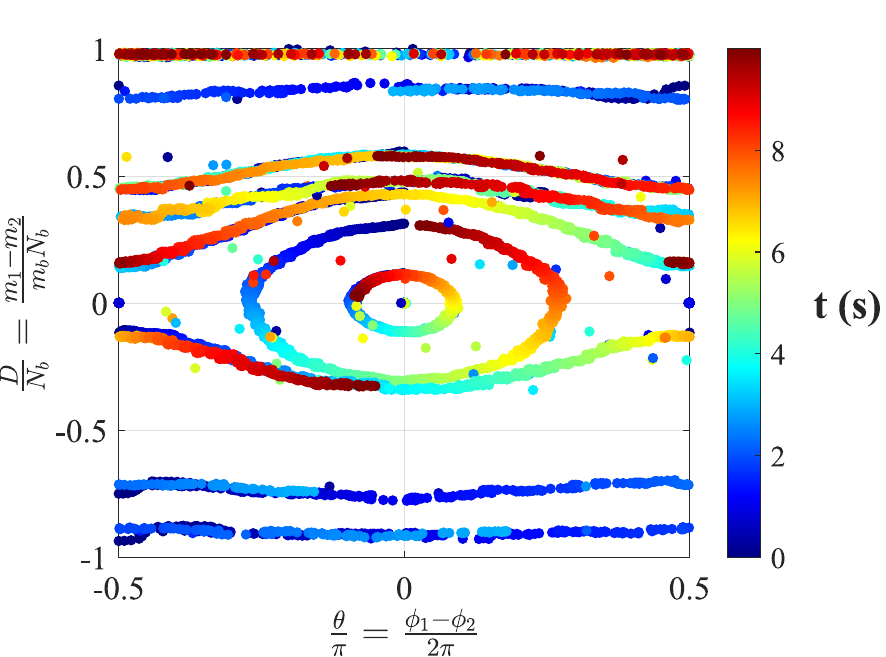}
    \caption{ GPE $(\theta, D)$ curves for a system with $N_b= 2000$ and $r_1/R\simeq 0.31$. Two fixed points of the BJJ model are visible along the $D=0$ axis.}
    \label{fig:pp_h}
\end{figure}

\begin{figure}
    \centering
    \includegraphics[width=0.5\textwidth]{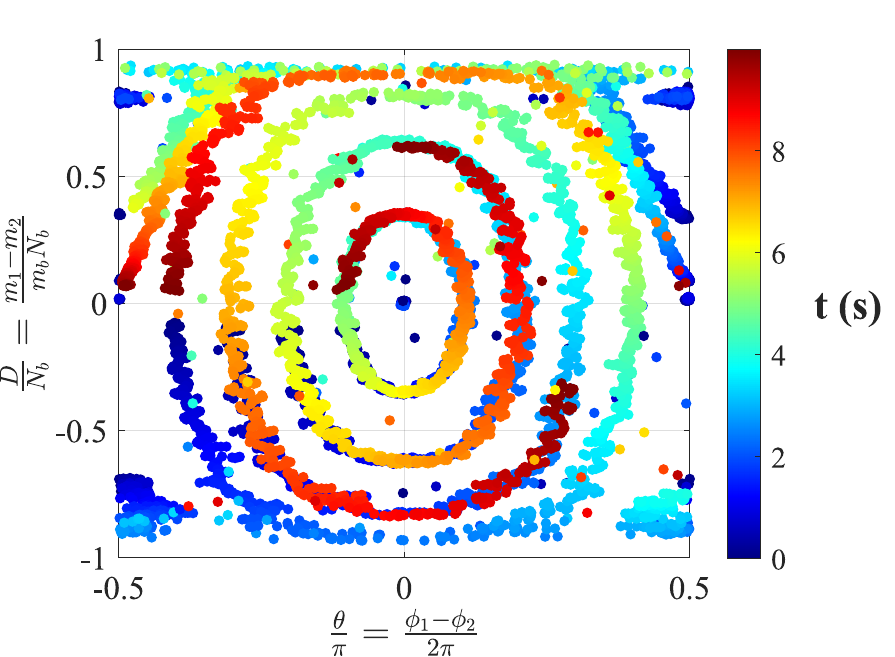}
    \caption{ GPE $(\theta, D)$ trajectories for a system with $N_b= 450$ and $r_1/R\simeq 0.22$. The numerics reproduce the four characteristic fixed points featured by a BJJ in this regime: two maxima, one saddle point and one minimum. }
    \label{fig:pp_aST}
\end{figure}

Remarkably, both the linear region of the phase portrait as well as the nonlinear phenomena, such as the macroscopic quantum self-trapping, are captured. 
We extract the $(\theta, D)$-trajectories, where $\theta$ is half the $b$-peaks phase difference, $\theta=\frac{\phi_1-\phi_2}{2}$, and $D$ is the peaks mass imbalance.
The curves, at different initial conditions, resemble the iso-energetic levels of the BJJ phase-portrait. 
Figs. \ref{fig:pp_d}-\ref{fig:pp_aST} show at a mean-field level that the co-rotating vortex pair makes up a Bosonic Josephson Junction.
Here, the oscillations of the phase difference of the two local peaks are coupled with their mass imbalance.
This phenomenon is possible thanks to the existence of circular orbit solutions for the vortex pair, that support a mass flux of the $b$-component in the effective double-well potential given by the vortex wells. As mentioned, we interpret in the first place the two vortices as a time-independent potential, in a rotating frame of reference.
In this way, the BH-dimer dynamics is effectively decoupled from the vortex pair dynamics.

Figs. \ref{fig:pp_d}-\ref{fig:pp_aST} show examples of $(\theta, D)$ trajectories for different systems, corresponding to different structures of the phase-space in the BH formalism. The characteristic BJJ trajectories are present, stable over time, and robust. The initial conditions on $\theta$ and $D$ do not compromise the BJJ dynamics. Not only the characteristic plasma and $\pi$ oscillations are present, but also, remarkably, the separatrices of such domains are well reproduced by the numerics.
The method we employ for extracting $\theta$ is subject to small artifacts (outliers) that are however not of concern as we neglect them in our results' analysis.
Although we did not plot all the trajectories, the phase portraits are symmetric with respect to the origin so that the self-trapping can involve the large population of any of the two vortices. Note that the angle $\theta$ is naturally endowed with periodic boundary conditions, leading to a
the cylinder geometry of the phase space.
Moreover, the BH-dimer dynamics frequency trend in the plasma oscillations region,
well visible in Figs. \ref{fig:pp_d} and \ref{fig:pp_aST}: the larger the circular orbits, the smaller is the frequency.

Fig. \ref{fig:pp_d} shows an example where both the plasma (around the origin) and the $\pi$ oscillations (around $|\theta|=0.5\;\pi$) of the peaks' population imbalance $D$ and their local phase difference $\theta$ are present.
$D$ and $\theta$ are extracted as described in section \ref{app:code_description}. One can see that the 
inter-vortex tunneling
allows for Josephson oscillations of different amplitude around the origin and that their frequency is smaller the larger they are. In the origin we see that a fixed point is reproduced by the GPE (the minimum of energy for a two-site BH model), while at large phase difference we see orbits characterized by large $D$ variations, while $|\theta|$ is within some interval around $0.5\;\pi$. In $|\theta|=0.5\;\pi$ and $D=0$, two other fixed points are correctly found. These correspond to maxima of the energy in a BH dimer.
Around the separatrix between the domain including the origin and the higher $|\theta|$ domains, the numerical noise, as expected, increases. Remarkably, the ordered 
BH-dimer-like dynamics emerges as well in the highly nonlinear, far-from-the-origin, regions. Here, high values of the phase difference are involved, as well as large fluctuations of the mass imbalance. Also, note that in the limit $D=\pm 1$ one of the two vortices has almost zero mass.
\\The system in Fig. \ref{fig:pp_h} represents a BJJ in another regime: here the on-site interactions are strong enough so to show self-trapping, i.e. the $D$-oscillations around a mean value different from zero. In this case, these are coupled to the ballistic evolution of the phase difference, which reproduce in some cases the \textit{ac} Josephson effect.
The ballistic orbits are characterized by large variations of $\theta$, while $|D|$ stays confined around a mean value. Two fixed points of the BJJ model are clearly visible along the $D=0$ axis: The minimum in the origin and the saddle point at $|\theta|=0.5\;\pi$.
The vortex pair contains here a larger number of particles $N_b$ and the vortex centers are more distanced with respect to the case of Fig. \ref{fig:pp_d}. 
\\Finally, the BJJ of Fig. \ref{fig:pp_aST} features a regime where the macroscopic quantum self-trapping occurs around $|\theta|=0.5\;\pi$.
 Such trajectories describe a case where one of the two co-rotating vortices has a large mass, while the other have a very small mass, so that the $b$-population is always in its largest part trapped within a specific vortex. The self-trapping regions present large values of $D$ and occur at large values of $|\theta|$, so that both quantities vary within a relatively small interval. The numerics reproduce the four characteristic fixed points featured by a BJJ in this regime: two maxima at large $|D|$ values and $|\theta|=0.5\;\pi$ (note the periodic boundary conditions in $\theta$), a saddle point in $|\theta|=0.5\;\pi$ and $D=0$, and the minimum in the origin.

\section{Theoretical modelling: estimate of the BH parameters}
\label{sec:theoretical_modelling}

Given a circular-orbit solution \cite{Bellettini2024}, where the two vortices are rotating specularly to each other around the origin at frequency $\Omega$, we can treat the vortex pair as an effective external potential for the component $b$. When switching to a rotating reference frame of frequency $\Omega$, the effective external potential becomes virtually time-independent.
Hence, the Hamiltonian relative to the component $b$, in the second quantization formalism, is

\begin{equation}
    \mathcal{H}= \mathcal{H}_0 + \mathcal{U}= \mathcal{H}_0 + \int d^3 r\; \frac{g_b}{2} (\hat{\psi}_b^+)^2\hat{\psi}_b^2,
    \label{eq:Ham_sec_quant}
\end{equation}

\begin{equation}
\begin{split}
    \mathcal{H}_0 &= \int d^3 r\; \left( -\frac{\hbar^2}{2 m_b} \hat{\psi}_b^+ \Delta\hat{\psi}_b\right) +\\
    &+g_{ab}\int d^3 r\; \;\rho_a\hat{\psi}_b^+ \hat{\psi}_b-\Omega \int d^3 r\; \hat{\psi}_b^+ L_3 \hat{\psi}_b 
\end{split}
\label{eq:hzero}
\end{equation}

and $L_3$ is the z-component of the angular momentum

\begin{equation}
    L_3=-i\hbar \left( x\frac{\partial}{\partial y} - y \frac{\partial}{\partial x} \right).
\end{equation}

 In Appendix \ref{sec:ang_mom_integral}, we show that for two co-rotating vortices the angular momentum term, in Hamiltonian \eqref{eq:hzero}, is zero.
In the above, $\rho_a =\rho_a(\bm r) = |\psi_a(\bm r)|^2$ is the density profile of the majority component $a$ and all the spacial vectors are 2D. 
Note that the component $a$ only contributes to $\mathcal{H}$ via the term $g_{ab}\rho_a$, which is the effective external potential felt by the component $b$ and is treated as time-independent.

From the Gaussian \textit{Ansatz} in Eq. \eqref{eq:rho_a} \cite{Bellettini2023} we derive the oscillation frequency $\omega$ of any of the wells in the locally harmonic potential approximation.
As aforementioned, we consider here identical wells at distance $r_1$ from the origin. Hence, the two wells have the same Gaussian width $\sigma_a$.

\begin{equation}
    \omega^2 =\frac{2 g_{ab} n_a}{ m_b \sigma_{a}^2 L_z}, \;\;\;  n_a =\frac{N_a}{\pi(R^2-2\sigma_a^2)}.
\end{equation}

\begin{equation} 
\rho_a=\frac{n_a }{L_z}\bigg(  1- \sum_{i=1}^{2} e^{-\frac{|\bm{r}-\bm{r}_{v,i}|^2}{\sigma_{a}^2}}  \bigg),
\label{eq:rho_a}
\end{equation}
where $\bm{r}_{v,i}$ is the 2D vector of coordinates of vortex $i$.

\subsection{Space-mode approximation}

Starting from the Hamiltonian \eqref{eq:Ham_sec_quant}, we perform a two-mode approximation in the spirit of Ref. \cite{Milburn1997}.
We expand the field operator $\hat{\psi}_b$ in terms of space modes

\begin{equation}
    \hat{\psi}_b  = \sum_{i=1}^{2} \hat{b}_i(t) W_i(\bm{r}), \;\;\; i=1,2    \label{eq:two_mode_expansion}
\end{equation}
where $\hat{b}_i$ are the bosonic mode operators, such that $[\hat{b}_i,\hat{b}_j]$ $=[\hat{b}^+_i,\hat{b}^+_j]=0$ and $[\hat{b}_i,\hat{b}^+_j]=\delta_{ij}$.
The functions $W_i(\bm{r})$, in Eq. \eqref{eq:two_mode_expansion}, are the ground state wavefunctions of the right well ``$1$" and of the left well ``$2$" (corresponding to the two effective potential wells 
placed at the vortex cores in component $a$), in the local harmonic approximation 
of the well profiles

$$ 
V_{har} = \frac{1}{2} m_b \omega^2 |\bm{r}-\bm{r}_{v,i}|^2,
\quad 
\omega^2
= g_{ab} \frac{2n_a}{m_b L_z \sigma_{a}^2}.
$$

From now on, the dependency of the mode operators on time and of $W_i$ on space are left implicit. Function $W_i$
are given by

\begin{equation}
    W_i  = \gamma 
    \exp \Bigl (
    -\alpha |\bm{r}-\bm{r}_{v,i}|^2 \Bigr ) ,\quad
    \alpha = \frac{m_b\omega}{2\hbar},
\label{eq:space_mode_apprx}
\end{equation}

with the normalization constant
$\gamma =  \sqrt{{2\alpha}/{(\pi L_z)}}$.
Note that at large values of ${g_{ab}}/{\sqrt{g_a g_b}}$ different functions $W_i$ might be more appropriate to precisely capture the BJJ dynamics.

After plugging the \textit{Ansatz} described by  Eq. \eqref{eq:two_mode_expansion} into the Hamiltonian \eqref{eq:Ham_sec_quant}, we obtain the
final $b$-Hamiltonian in the two-mode approximation 
(refer to Appendix \ref{sec:two_mode} for its explicit derivation)

\begin{equation}
\begin{split}   
    \mathcal{H}_{tm} &=
    \frac{U}{2}  \sum_{i=1}^{2} \;\hat{n}_i(\hat{n}_i-1)-\mu N_b - J\; (\hat{b}^+_{2}\hat{b}_1 -  \hat{b}^+_{1}\hat{b}_2) 
\end{split}    
\label{eq:htm}  
\end{equation}

with $N_b = \sum_{i=1}^{2} \;\hat{n}_i $,
$\hat{n}_i=\hat{b}_i^+\hat{b}_i$ the particle number operator of site $i$, and
$$
\mu = -\frac{g_{ab}n_a}{L_z(1+2\alpha\sigma_a^2 )},
$$
the chemical potential. The contribution $\mu N_b$ can be neglected as $N_b$, being a constant of motion, does not affect the properties of the system. The interaction and tunneling parameters,
\begin{equation}
%
U=
\frac{g_b\alpha}{\pi L_z}=
\frac{g_b}{\hbar\pi L_z}
\sqrt{\frac{g_{ab}\;n_a m_b}{2\sigma^2_{a}L_z}}    \label{eq:U}
\end{equation}
and
\begin{equation}
    \begin{split}
        J= &\;\frac{g_{ab}n_a}{L_z}\bigg[\frac{1}{\sigma_a^2 }\left(\frac{\hbar}{m_b\omega}+r_1^2\right) e^{-\frac{m_b\omega}{\hbar}r_1^2}+
        \\
        &+\frac{4}{2+1/(\alpha\sigma_a^2)}  e^{
        - \frac{4\alpha r_1^2 (1+\alpha \sigma_a^2)}{1+2\sigma_a^2}}\bigg]
    \end{split}
    \label{eq:J}
\end{equation}
respectively, are calculated for a double-well system corresponding to a vortex pair where the vortices are at the same distance from the disk center but in opposite positions (${\bf r}_2 = -{\bf r}_1$).
Note the dependency, in Eqs. (\ref{eq:U}) and (\ref{eq:J}), on the macroscopic vortex parameters. At increasing $g_{ab}n_a$ the harmonic-oscillator (squared) frequency $\omega^2$ increases, leading in turn to a stronger confinement and hence to a rise in the onsite interaction $U$. 
Conversely, $U$ decreases with increasing $\sigma_a$, i.e. the Gaussian vortex-profile width. 
Furthermore, $U$ linearly depends on the $b$-atoms interaction parameter $g_b$, as in standard BH models, while
the tunneling parameter $J$ rapidly decreases at increasing $r_1$, i.e. the vortices' half-distance.
It is also worth observing that
the second term in Eq. \eqref{eq:J} represents an extra contribution with respect to the standard tunneling parameter of Ref. \cite{Milburn1997}.
This is due to the fact that the effective potential created by the vortices features a harmonic characteristic length $(2\alpha)^{-1/2}$ that is comparable with the vortex width $\sigma_a$ (of the same order of the component-$a$
healing length), and to $\sigma_a$
being in general not negligible with respect to the vortices' distance $2\,r_1$.

In Appendix \ref{sec:two_mode} we derive the general expression of $J$ for the case where the two vortices rotate onto two different circumferences. We proved the existence of these configurations for two individual massive vortices in Ref. \cite{Bellettini2024}, and their extension to a BJJ is analogous to the single-orbit case.

Note that according to Milburn, Corney \textit{et al.} \cite{Milburn1997} the two-mode approximation is only valid in the limit

\begin{equation}
    N_b \ll \sqrt{\frac{\hbar}{2m_b\omega}}\frac{1}{a_b}=N_{h},
    \label{eq:limit_of_validity}
\end{equation}
i.e. when the intraspecies interactions do not perturb significantly the ground state of the harmonic oscillators. Note as well that, as $N_h\propto 1/(a_b \sqrt[4]{g_{ab}})$, the smaller is the scattering length of $a_b$ or the smaller is $g_{ab}$, the more the \textit{Ansatz} \eqref{eq:two_mode_expansion} is appropriate for capturing the double-well phenomenology

\subsection{Mean-field approximation}

Since the numbers of bosons contained in the two wells is rather large, the purely quantum description based on Hamiltonian \eqref{eq:htm} can be replaced by the mean-field BH model obtained within the coherent-state picture of Ref. \cite{FRANZOSI2000}. The mean-filed Hamiltonian is given by 
$\mathcal{H}_{mf} = \langle \psi_b| \mathcal{H}_{tm}|\psi_b\rangle$ where $|\psi_b\rangle$ reads
\begin{equation*}
    |\psi_b\rangle=e^{i \frac{S}{\hbar}}|Z\rangle, \;\;\; |Z\rangle=\otimes_{i=1,2}|z_i\rangle
\end{equation*}
and $S$ is the effective action related to Hamiltonian $H_{mf}$. State
$|z_i\rangle$ is the Glauber coherent state whose parameter 
$z_i= \langle z_i | b_i |z_i\rangle 
\in \mathbb{C}$ represents the local order parameter of the well $i$ and allows one to define the relevant average boson population $|z_i|^2= \langle z_i | b^+_i b_i |z_i\rangle$. The resulting mean-field Hamiltonian is

\begin{equation*}
    H_{mf}=\sum_{j=1}^2 \frac{U}{2}|z_j|^4-
    J\bigg(z_2^* z_1+z_1^* z_2\bigg)
\end{equation*}
whose derivation is discussed in Appendix \ref{sec:mean_field}.
With a convenient change of variables, the phase space is reduced to two dimensions. 
By defining the new pairs of canonically conjugate variables
$$
D=|z_1|^2-|z_2|^2,\,\, \theta=\frac{\phi_1-\phi_2}{2},
$$
$$
\mathcal{N}=|z_1|^2+|z_2|^2, \,\, \psi=\frac{\phi_1+\phi_2}{2}, 
$$
where $z_j=|z_j|e^{i\phi_j}$, the following final BJJ Hamiltonian 
\begin{equation}
    H_{mf}=\frac{U}{4}\mathcal{N}^2+\frac{U}{4}D^2-J\sqrt{\mathcal{N}^2-D^2}\cos{(2\theta)}
    \label{eq:Heff_final}
\end{equation}
is obtained.
Note that the variable $\mathcal{N}$ is a constant of motion, representing the total number of particles $N_b$ of the BJJ, and $\psi$ is an auxiliary variable, so that the phase space is made up by the two conjugate variables $D$ and $\theta$, where
$D$ represents the population imbalance between the two wells, while $\theta$ their phase shift.
The relevant equations of motion read (the associated Poisson Brackets are given in Appendix \ref{sec:mean_field})
\begin{align}
    & \hbar\dot{D}=2J\sqrt{\mathcal{N}^2-D^2}\sin{(2\theta)}\label{eq:Ddot} \\
    & \hbar\dot{\theta}=-\frac{U}{2}D-\frac{JD}{\sqrt{\mathcal{N}^2-D^2}}\cos{(2\theta)}\label{eq:thetadot}
%
\end{align}
while $\dot{\mathcal{N}}=0$ and the equation for $\psi$ plays the role of an auxiliary equation.
Again, following Ref. \cite{FRANZOSI2000} it is convenient to define the quantity 
$$\Gamma
=\frac{2J}{U\mathcal{N}},$$
discriminating different regimes of the phase portrait.
Finally, the frequency $\nu$ of the $(\theta,D)$ orbits in the small-oscillations limit [i.e. close to $(0,0)$]
is
\begin{equation} 
    \nu =\sqrt{\frac{2J\mathcal{N}U}{\hbar^2}\left(1+\frac{2J}{\mathcal{N}U}\right)}= 
     2J\sqrt{\frac{1}{\Gamma}(1+\Gamma)},
    \label{eq:freq_small_osc} 
\end{equation}
while a solution of Eqs. \eqref{eq:Ddot}-\eqref{eq:thetadot} in the linear domain is given by $D=A\cos{(\nu t)}$
and $\theta=-\frac{A\nu\hbar}{4J\mathcal{N}}\sin{(\nu t)}$ with $A$ depending on the initial condition.

\section{Results and discussion: comparison with the GPE dynamics}
\label{sec:comparison}

We compare the GPE results with the two-well mean-field BH model predictions, using for the BH parameters the analytical expressions 
\eqref{eq:U} and
\eqref{eq:J} in terms of the parameters of the vortex pair. 
We take a series of 
parameter pairs $(U,J)$ while only $N_b$ varies and we compare the GPE phase portraits and the real-time dynamics with the trajectories obtained via our model. Here, we set  $\sigma_a = 0.179\;R$. $N_a =10^5$ and $r_1\simeq 0.22\; R$.

\begin{figure}
    \centering
    \includegraphics[width=0.5\textwidth]{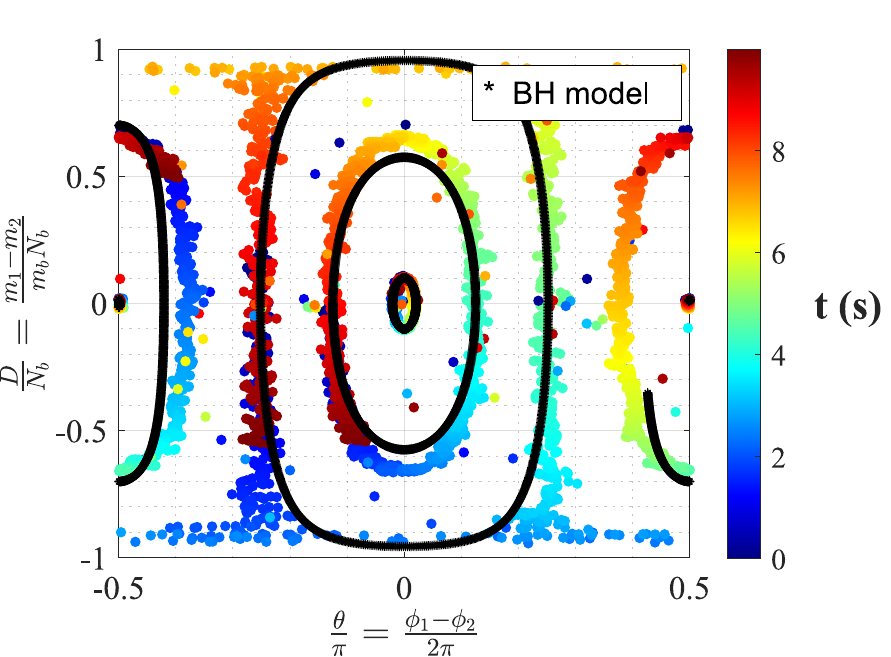}
    \caption{Comparison between the GPE-extracted phase portrait and our analytical BH model. $N_b=10$, $\Gamma\simeq 1.5$.}
    \label{fig:pp_p_comp}
\end{figure}

\begin{figure}
    \centering
    \includegraphics[width=0.45\textwidth]{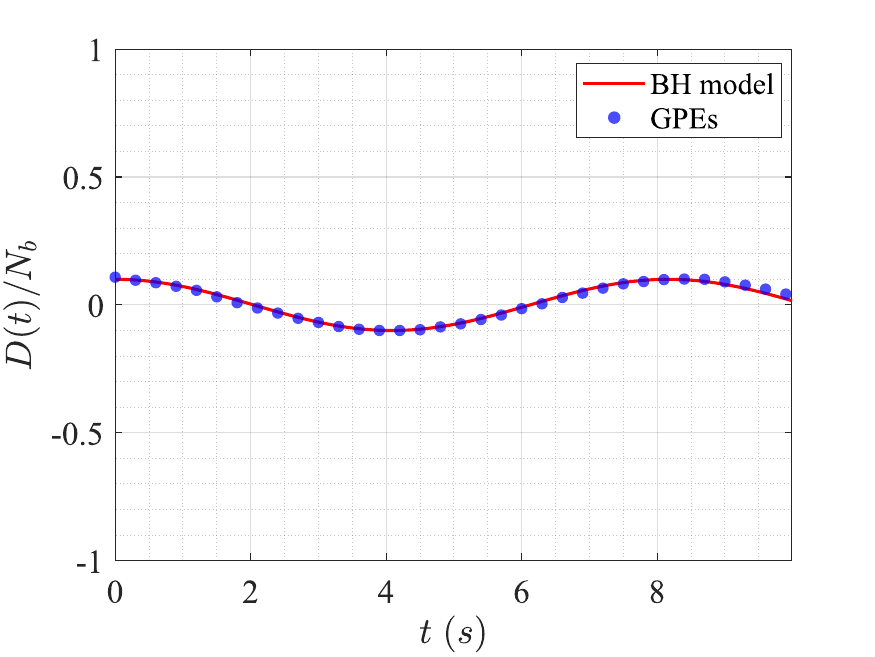}
    \caption{Comparison between the time evolution of the population imbalance $D$ as extracted from the GPE and as predicted by our analytical model, for a trajectory close to the origin in the phase space, i.e. in the linear domain. The system is the same of Fig. \ref{fig:pp_p_comp}, $N_b=10$.}
    \label{fig:D_p5_comp}
\end{figure}

\begin{figure}
    \centering
    \includegraphics[width=0.45\textwidth]{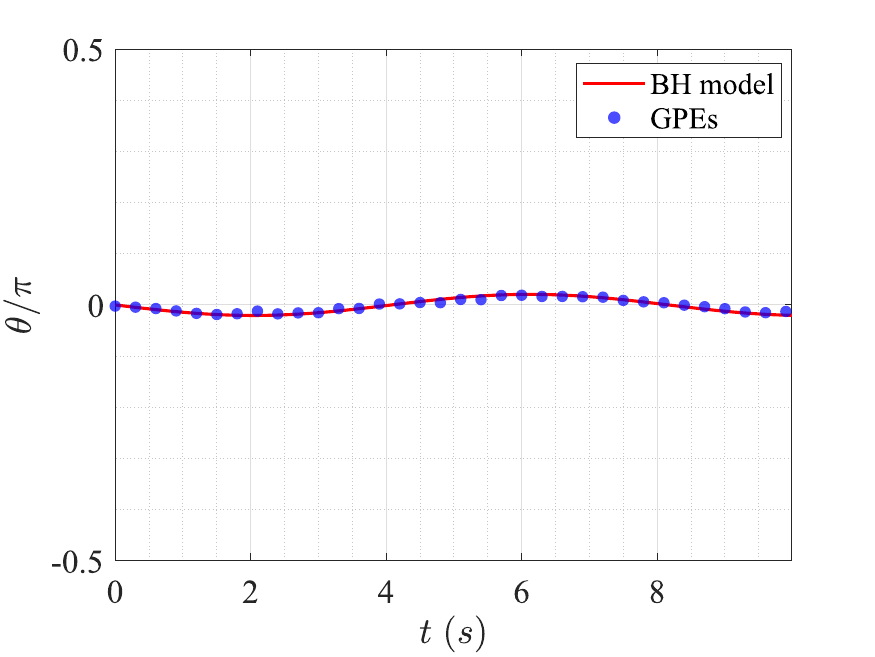}
    \caption{Comparison between the time evolution of the phase shift $\theta$ as extracted from the GPE and as predicted by our analytical model, for the same trajectory of Fig. \ref{fig:D_p5_comp}.}
    \label{fig:theta_p5_comp}
\end{figure}

\begin{figure}
    \centering
    \includegraphics[width=0.45\textwidth]{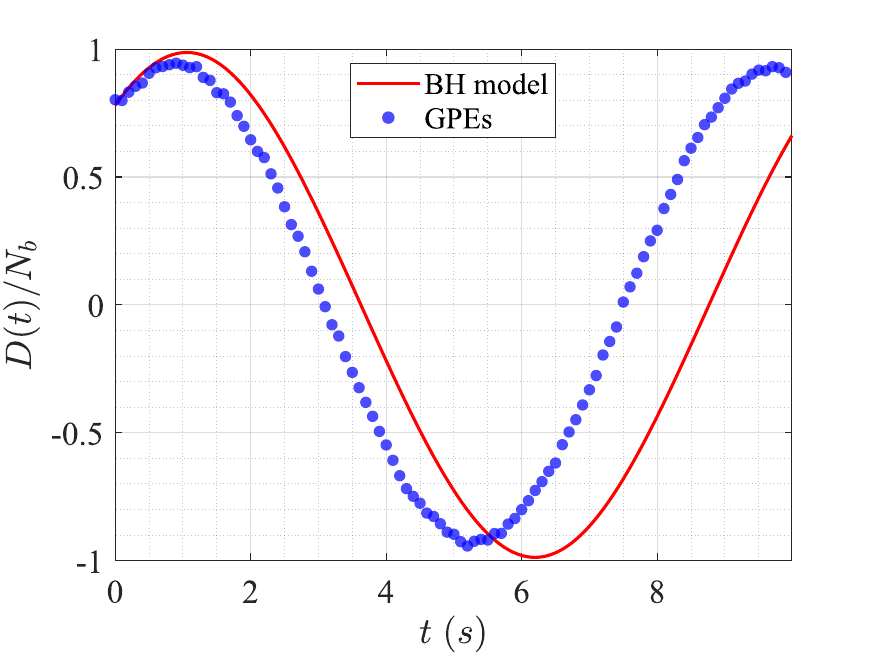}
    \caption{Same as Fig. \ref{fig:D_p5_comp} for a trajectory in the strongly nonlinear domain.}
    \label{fig:D_p1_comp}
\end{figure}

\begin{figure}
    \centering
    \includegraphics[width=0.45\textwidth]{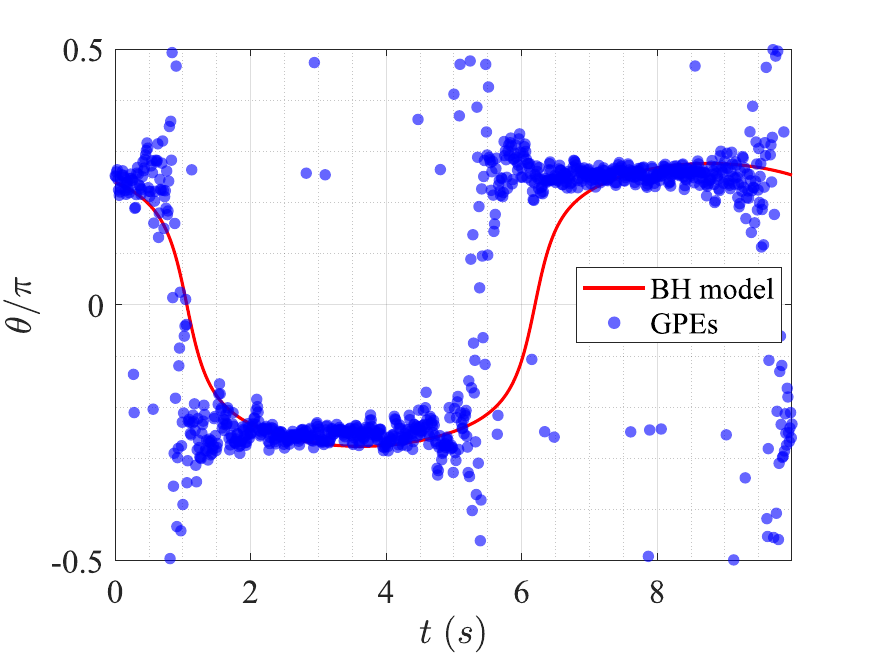}
    \caption{Same as Fig. \ref{fig:theta_p5_comp} for the same trajectory of Fig. \ref{fig:D_p1_comp}.}
    \label{fig:theta_p1_comp}
\end{figure}

We find that for a small number of particles, i.e. $N_b=10$,   
there is a general good agreement between the model and the GPE, both in the phase portrait and in the BH-dimer dynamics. As visible in Fig. \ref{fig:pp_p_comp}, a mismatch stems in the strongly nonlinear regime, characterized by $\pi$-phase oscillations ($\theta (t) \approx \pi/2$). These are however still qualitatively captured. 
Besides, Figs. \ref{fig:D_p5_comp}-\ref{fig:theta_p1_comp} show the comparison of the GPE and BH dynamics, firstly in the plasma oscillations domain (which features small oscillations of $D(t)$ and $\theta (t)$) around zero) and secondly for a large orbit at the boundary with the $\pi$-oscillations domain. Here we see that not only the shape of the trajectories in the phase space are in general well captured, but also the BH-dimer dynamics is, at least in the plasma oscillations domain. When going further from the linear region, our model underestimates the frequency of BH-dimer dynamics. Here the ratio of the GPE-extracted frequency to the BH model frequency $\nu_{GPE}/\nu_{BH}$ can vary approximately from $1.1$ to $1.6$ for the tested trajectories.

To sum up, in the small-$N_b$ limit, the analytical expressions for the BH model parameters are reliable and the macroscopic effects occurring in the BJJ are well described. For trajectories in the phase space that are further from the plasma-oscillation domain, some small discrepancy arises (Figs. \ref{fig:D_p1_comp}-\ref{fig:theta_p1_comp}). These involve situations of a larger mass-imbalance between the two vortices, which may excite other phenomena in the GPE simulation. In general, this suggests that this discrepancy is not of concern, rather, the still persisting good agreement is remarkable.



\begin{figure}
    \centering
    \includegraphics[width=0.5\textwidth]{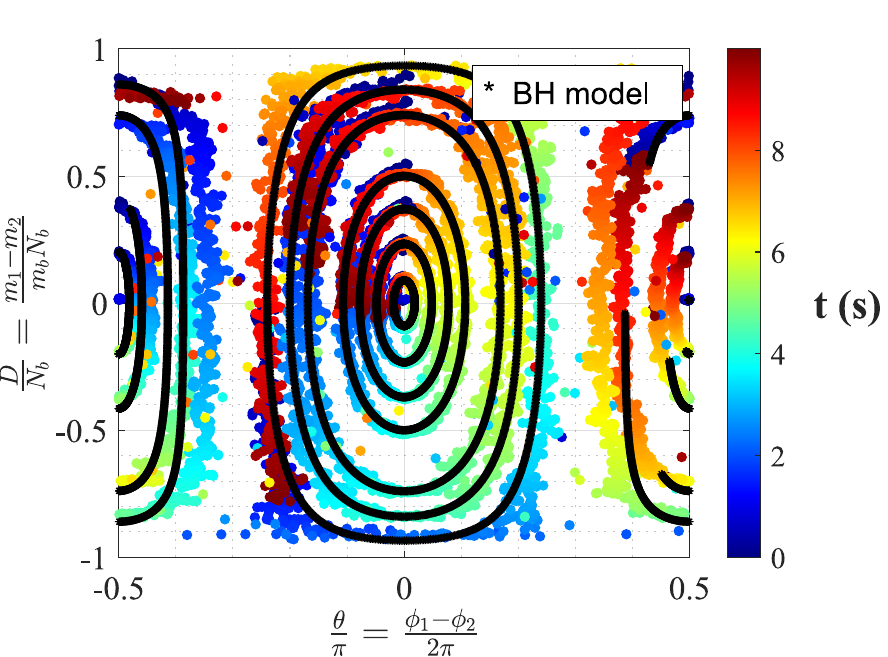}
    \caption{Comparison between the GPE-extracted phase portrait and our analytical BH model. $N_b=50$, $U_{eff}=U/5$, $\Gamma\simeq 1.5$.}
    \label{fig:pp_o_comp}
\end{figure}

\begin{figure}
    \centering
    \includegraphics[width=0.5\textwidth]{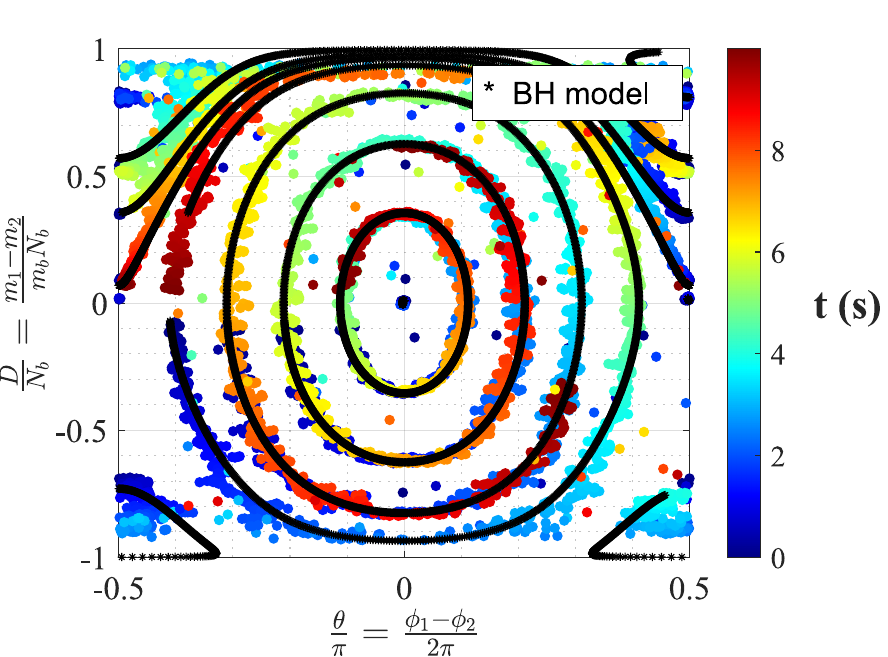}
    \caption{Comparison between the GPE-extracted phase portrait and our analytical BH model. $N_b=450$, $U_{eff}=U/11$, $\Gamma\simeq0.4$. }
    \label{fig:pp_aST_comp}
\end{figure}

\begin{figure}
    \centering
    \includegraphics[width=0.5\textwidth]{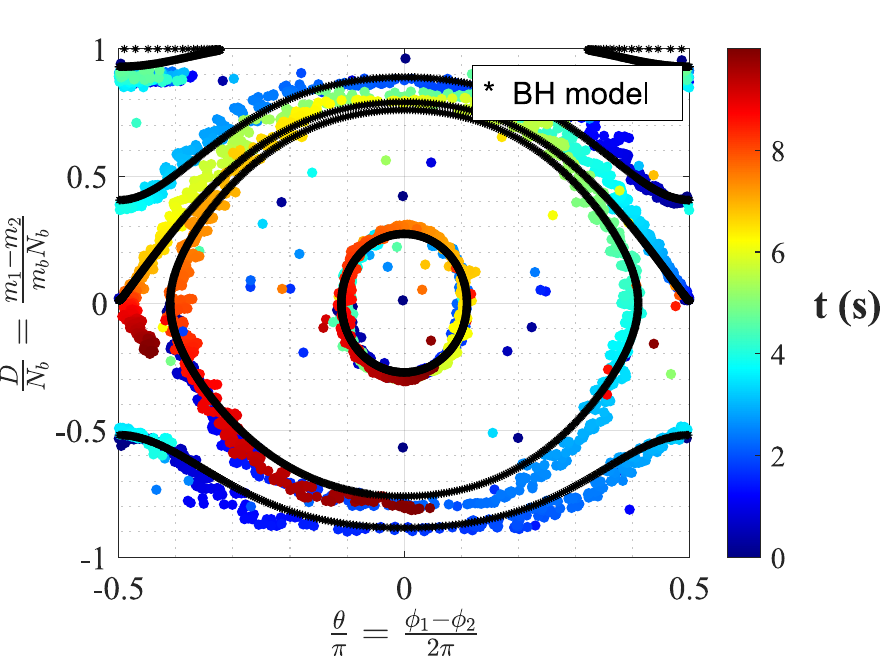}
    \caption{Comparison between the GPE-extracted phase portrait and our analytical BH model. $N_b=1000$, $U_{eff}=U/13$, $\Gamma\simeq 0.2$. }
    \label{fig:pp_a_comp}
\end{figure}

\begin{figure}
    \centering
    \includegraphics[width=0.45\textwidth]{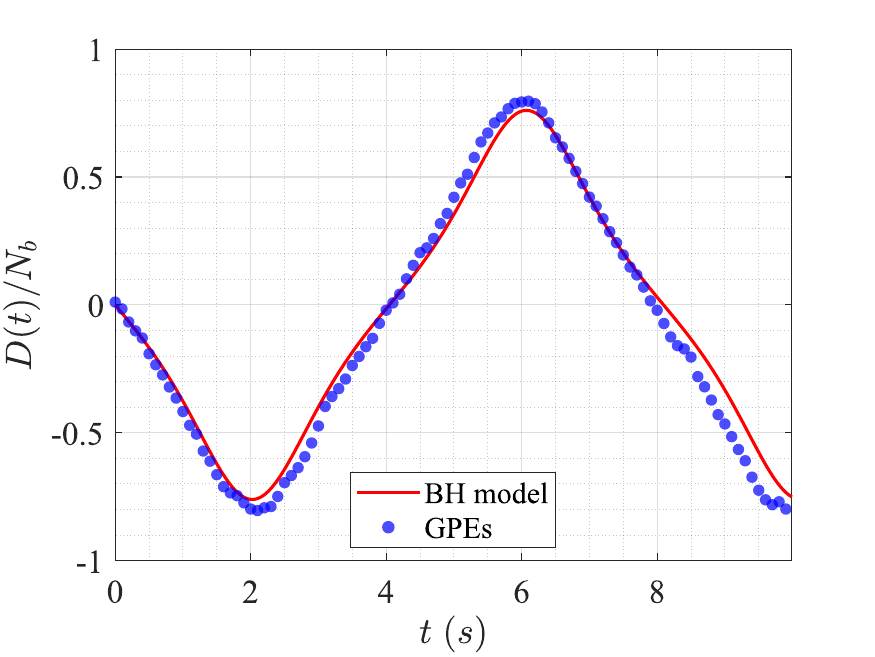}
    \caption{Comparison between the time evolution of the population imbalance $D$ as extracted from the GPE and as predicted by our analytical model, for a trajectory close to the separatrix in the phase space, between the plasma oscillations region and the ballistic phase shift oscillations. The system is the same of Fig. \ref{fig:pp_a_comp}, $N_b=1000$, $U_{eff}=U/13$. }
    \label{fig:D_a5_comp}
\end{figure}

\begin{figure}
    \centering
    \includegraphics[width=0.45\textwidth]{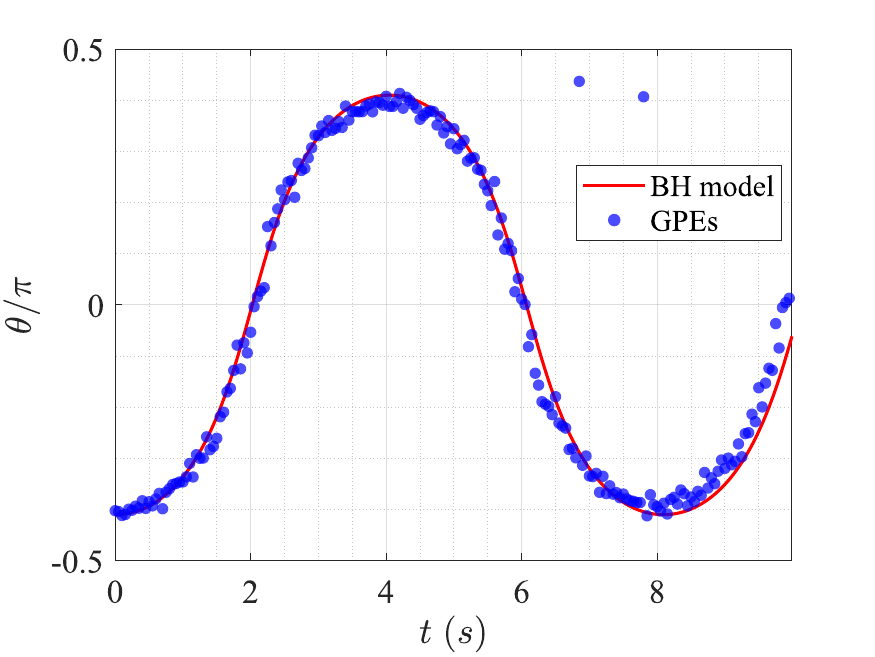}
    \caption{Comparison between the time evolution of the phase shift $\theta$ as extracted from the GPE and as predicted by our analytical model, for the same trajectory of Fig. \ref{fig:D_a5_comp}. }
    \label{fig:theta_a5_comp}
\end{figure}

The larger $N_b$, the more evident arises a mismatch, which is however substantially improved via an effective correction of the coefficient $U$ [equation \eqref{eq:U}].
In this case, we obtain an overall good agreement of the phase portraits (Figs. \ref{fig:pp_o_comp}, \ref{fig:pp_aST_comp}, and \ref{fig:pp_a_comp}), where some not qualitative discrepancies again arise in the strongly nonlinear domain. 
In general, while the double-well dynamics is often well reproduced by the numerical simulations, which show a good agreement with the trajectories of the BH phase portrait,
the frequency (affecting the evolution of $D$ and $\theta$) can be underestimated by our model of a factor $\nu_{GPE}/\nu_{BH}$ that goes approximately from $1.1$ to $1.6$ for the examined trajectories.
The examples of Figs. \ref{fig:D_a5_comp}-\ref{fig:theta_a5_comp} are cases of a very good agreement between the frequency of the BH and the GPE-extracted dynamics. 

The effective correction of $U$ is such that $U_{eff}=U/f$, where $f>1$ and in most cases $f$ increases with $N_b$. We interpret this correction as an effective correction of $g_b$, namely as an effect of the intraspecies interactions, which are not taken into account in the space-mode approximation. 
The \textit{Ansatz} \eqref{eq:space_mode_apprx} is in fact valid in the limit expressed by Eq. \eqref{eq:limit_of_validity}.
Parameter $g_b$ controls the repulsive interactions, resulting in a wider $b$-peak with respect to the noninteracting case. Furthermore, this effect is larger, the larger $N_b$, something that gives a reasonable interpretation of the $U$ correction. Possible ways to improve the model include the numerical computation of the BH parameters, or models that go beyond the standard two-mode approximation \cite{Ananikian2006, Burchianti2017}.

Finally we mention an alternative scheme that could be adopted to interpret the discrepancies. 
One could in fact always leave $U$ untouched and tune the parameter $\sigma_a$ for every system, at most to large effective values, 
so that the phase portrait provided by numerical simulations 
features a very good agreement with that of the  double-well BH model. This, however, at the price of losing
the agreement in the real-time dynamics, obtaining in fact a large overestimation of the frequency by the model, which increases at increasing $N_b$.


\section{Conclusions and outlook}
\label{sec:conclusion}

To sum up,
the thorough numerical study of the dynamics of a massive-vortex pair unveiled
a robust realization of the BJJ, triggered by the inter-vortex $b$-bosons tunneling.
We extracted several phase portraits from the simulation of the coupled Gross-Pitaevskii equations and observed, remarkably, all the characteristic phenomena featured by a BJJ, including the plasma and $\pi$ oscillations and the macroscopic quantum self-trapping. Thanks to the existence of circular orbit solutions for the two vortices \cite{Bellettini2024}, we were able to treat the vortex pair as fixed point solution in a rotating frame of reference. Thus, they constitute a time-independent effective potential for their in-filling component, tunneling within the two-vortex effective double-well potential. Despite the small, fast variations of the vortex radial positions, the BH-dimer dynamics stayed unaffected and stable over time. Hence, vortices in BECs can be used in replacement to optical potentials to support the formation of a BJJ.
The remarkable fact that vortices can host a component undergoing quantum tunneling 
offers a rich and intriguing scenario. This suggests future extensions starting from the minimum building block, that is the vortex pair, here presented.

We modeled the BJJ Hamiltonian associated to the component $b$ 
filling the two vortex cores. In 
our system, $\psi_a$ represents a background time-independent order parameter. Starting from the quantum $b$-Hamiltonian,
we performed a two-mode approximation and then derived, via a coherent state approach, its corresponding mean-field version. We presented a final Bose-Hubbard model where the parameters are explicitly expressed in terms of the system's physical quantities. In the derivations, we considered $b$ as localized around the centre of two locally harmonic wells, a condition justified by the 
immiscibility of the two species.

Finally, we compared our analytic Bose-Hubbard model with the numerical results. We found a very good agreement of the model with the GPE in the small-$N_b$ limit, both in the phase portraits and in the BH-dimer dynamics, while at larger $b$-particles numbers an effective correction of the onsite interaction parameter $U$ is necessary to obtain a good agreement with the numerics. We interpreted this correction, increasing with $N_b$, as an effect of the intraspecies interactions, which lead to an enlargements of the in-filling for the two-mode approximation. Nonetheless, an eventual improvement of the model via a new \textit{Ansatz} for the modes $W_i$ is out of the scope of our paper, whereas it would be interesting for the future.
Noticeably, the effectively corrected model was shown to capture even the separatrices in the phase portraits and the strongly nonlinear phenomena qualitatively. Some quantitative discrepancies can arise in the nonlinear regions, where the BH-dimer frequency is underestimated of up to a factor $1.6$ for in the examined orbits. This is likely due to other phenomena occurring in the GPE, such as a more pronounced asymmetry in the vortex wells or larger variations of the vortex radial positions. 
As a future outlook, it would be in fact intriguing to incorporate into the model the coupling between the vortex dynamics and that of the BH-dimer, by taking into account the variations in the distance of vortices.

Other following up scenarios include the vortex BH trimer and the emergence of chaos \cite{Franzosi2003, Mossmann2006, Arwas2015}, the tunneling within a massive vortex necklace \cite{Hernández-Rajkov2024, Caldara2024}, and a lattice \cite{Prates2022}.
In addition, it would be interesting to model
the asymmetric BH dimer \cite{Pigneur2018, Gati2007}, where the two vortices have different sizes. This is made possible thank to the in-filling component influencing the healing length of the condensate $a$. 
Further interesting outlooks are the inclusion of multiply quantized vortices, whose stability is guaranteed by the in-filling component \cite{Richaud2023}, the case of attractive interactions \cite{Spagnolli2017}, and the tunneling of a BEC mixture \cite{Penna2017, Richaud2019}.

\section*{Acknowledgements}
A. R. received funding from the European Union’s
Horizon research and innovation programme under the Marie Skłodowska-Curie grant agreement \textit{Vortexons} (no.~101062887). A. R. further acknowledges support by the Spanish Ministerio de Ciencia e Innovaci\'on (MCIN/AEI/10.13039/501100011033,
grant PID2020-113565GB-C21), and by the Generalitat
de Catalunya (grant 2021 SGR 01411).
Computational resources were provided by HPC@POLITO (http://hpc.polito.it).

\begin{appendix}

\section{Numerical simulation}
\label{app:code_description}

After nucleating the vortices via the imaginary-time procedure, we start the “real-time” simulation. Here the dynamics of the two order parameters is simulated by resolving the GPEs through the fourth-order Runge-Kutta algorithm. 
We compute the two vortex masses by integrating the $b$-density field $\rho_b$ over two separated domains. These are obtained by dividing the disk domain (containing the mixture) by the axis of the segment through the two vortex centres. This method requires an initial estimate of the vortex positions, which is done via an image processing tool.
Finally, we compute the phases of $b$-peaks by integrating the $b$-phase field, weighted by $\rho_b$, onto separate domains that are further restricted in comparison with those employed for the masses. We restrict the data, in this case, to the neighbourhoods of the two peaks, setting to zero all the data outside the circles of radius $0.01\; R$ centred in the two peaks. This is done to minimize numerical artifacts. 
Note that the distance between the two vortices should be large enough for a meaningful extraction of the vortex masses and phases as described above.

\section{Space-mode approximation}
\label{sec:two_mode}

Here we present the derivation of the BH Hamiltonian in the general case of two vortices rotating on
different orbits of radii $r_1$ and $r_2$, always with an angular position shift of $\pi$.
Recall that we assume the vortex cores can be associated with a Gaussian width of $\sigma_{a}$, so that their corresponding
potential wells have a frequency $\omega$, in the harmonic approximation, given by
$$\omega^2 =\frac{2 g_{ab} n_a}{ m_b \sigma_{a}^2L_z}.
$$
Within the space-mode approximation the ground state of the $i$-the well associated to the boson mode $b_i$ is $W_i  = \gamma  e^{-\alpha  |\bm{r}-\bm{r}_{v,i}|^2}$, with $\alpha =\frac{m_b\omega }{2\hbar} $ and 
$\gamma =\sqrt{{2\alpha }/{(\pi L_z)}}$, 
and it holds $(W_i,W_j) =\delta_{ij}$. We assume $\sigma_{a }\ll R$ and,
justified by the immiscibility condition, $R^2 \gg {1}/{2\alpha }$. Hence, the $a$-density profile at the vortex sites and the $W_i$ wavefunctions are ``narrow Gaussians".

\subsection{Free Hamiltonian $\mathcal{H}_0$}

In the rotating reference frame the Hamiltonian $\mathcal{H} =
\mathcal{H}_0 +\mathcal{U}$
of the component $b$
exhibits two contributions. The second one, $\mathcal U$, describes boson-boson interaction and will be analyzed later. The first one 

\begin{equation*}
\begin{split}
    \mathcal{H}_0 &= \int d^3 r\; \left(-\frac{\hbar^2}{2 m_b} \hat{\psi}_b^+ \Delta\hat{\psi}_b\right)  
    +g_{ab}\int d^3 r\; 
    \rho_a ({\bf r})
 \hat{\psi}_b^+ \hat{\psi}_b
    \\
    &-\Omega \int d^3 r\; \hat{\psi}_b^+ L_3 \hat{\psi}_b = (I) + (II)
\end{split}    
\end{equation*}
where
$$
\rho_a ({\bf r}) = \frac{n_a}{L_z} \bigg(  1- \sum_{i=1}^{2} e^{-\frac{|\bm{r}-\bm{r}_{v,i}|^2}{\sigma_{a,i}^2}}  \bigg) 
$$
shows the $g_{ab}$-dependent coupling with the component-$a$ density and the angular-momentum term emerging from the rotation.
In the following we substitute the \textit{Ansatz} \eqref{eq:two_mode_expansion} into $\mathcal{H}_0$ and solve the integrals.

\subsection{Integral $(I)$}

\begin{align*}
    & (I)  =\int d^3 r\; \left(-\frac{\hbar^2}{2 m_b} \hat{\psi}_b^+ \Delta\hat{\psi}_b\right) +  g_{ab}\int d^3 r\; \rho_a \hat{\psi}_b^+ \hat{\psi}_b 
    \\
    &=\sum_{i, j=1}^{2}
    \hat{b}_i^+\hat{b}_j
    \! \int \! d^3 r\; W_i \bigg[ H_{ho} 
    +g_{ab}\rho_a 
    -  \frac{m_b \omega^2}{2}|\bm{r}-\bm{r}_{v,j}|^2 \bigg]W_j
    \\
    & = \sum_{i=1}^{2}\hat{b}_i^+\hat{b}_i E_{0} + \sum_{j,i =1}^{2}\hat{b}_i^+\hat{b}_j K_{ij}
\end{align*}
where the ground-state energy $E_0=\frac{\hbar \omega}{2}$ emerges from the action of the harmonic-oscillator (local)  Hamiltonian
$$ 
H_{ho}=
-\frac{\hbar^2}{2m_b}\Delta + \frac{m_b \omega^2}{2}|\bm{r}-\bm{r}_{v,j}|^2+
$$
on its ground state $W_j$, and
the two-index symbol $K_{ij}$ reads
$$
K_{ij} = \int d^3 r\; W_i \left[ g_{ab}\rho_a  -  \frac{m_b \omega^2}{2}|\bm{r}-\bm{r}_{v,j}|^2 \right] W_j.
$$ 
Hence, we can rewrite $(I)$ as 

\begin{align*}
    (I) &= \sum_{i=1}^{2} (E_{0} + K_{ii})\;\hat{b}^+_i\hat{b}_i+ \hat{b}^+_{2}\hat{b}_1\; K_{2,1} +   \hat{b}^+_{1}\hat{b}_2 \;K_{2,1} 
    \\
    &= \sum_{i=1}^{2} E_i\;\hat{n}_i+ \hat{b}^+_{2}\hat{b}_1\; K_{2,1} +   \hat{b}^+_{1}\hat{b}_2 \;K_{2,1} 
\end{align*}
After some long but not complex derivations, we obtain

\begin{equation}
    \begin{split}
        &K_{ij}=g_{ab}\frac{n_a}{L_z} \delta^2_{ij} -  \frac{\pi \gamma^2 g_{ab}n_a}{2\alpha+\frac{1}{\sigma_{a}^2}} 
        \sum_{l=1}^{2}e^{
        A_{\ell} (x_i, x_j)} e^{
        A_{\ell} (y_i, y_j)}
        \\
    &-\frac{\gamma^2}{2}  m_b\;\omega^2 L_z\;e^{-\frac{\alpha}{2}|\bm{r}_{v,j}-\bm{r}_{v,i}|^2} \frac{\pi}{4\alpha }\left(\frac{1}{\alpha }+ \frac{|\bm{r}_{v,j}-\bm{r}_{v,i}|^2}{2}\right)
    \end{split}
\end{equation}
where $\bm{r}_{v,i} = (x_i, y_i)$,
$$
A_{\ell} (x_i, x_j)= \exp\left [
\frac{\left(\alpha X_{ij} + \frac{x_{l}}{\sigma^2_{a}}\right)^2}{2\alpha+\frac{1}{\sigma_{a}^2}}
-
\alpha \left ( x_{i}^2 + x_{j}^2 \right ) - \frac{x_{l}^2}{\sigma_{a}^2}
\right ],
$$
$X_{ij}= x_{i} + x_{j}$, and $A_{\ell} (y_i, y_j)$ is found by replacing $x_i$  ($x_j$) with $y_i$ ($y_j$) in $A_\ell$.
With the substitution $x_1=r_1\cos{\Theta}$,
$y_1=r_1\sin{\Theta}$, $x_2=-r_2\cos{\Theta}$, $y_2=-r_2\sin{\Theta}$, i.e. taking two co-rotating vortices,
we get, for $i\neq j$

\begin{equation}
\begin{split}
    K_{ij}&= -\frac{g_{ab}n_a}{L_z}\bigg[ \frac{4}{2+1/(\alpha\sigma_a^2)}e^{-\frac{\alpha(1+\alpha\sigma_a^2)(r_1+r_2)^2}{1+2\alpha\sigma_a^2}}+\\
    &+\frac{e^{-\frac{\alpha}{2}(r_1+r_2)^2}}{2\sigma_a^2} \bigg(\frac{1}{ \alpha}+\frac{(r_1+r_2)^2}{2}\bigg)\bigg]
\end{split}    
\end{equation}

Finally, we approximate $K_{ii}$ by keeping in the summation over $l$ only the contribution of $l=i$, and obtain, for two co-rotating vortices
$$
K_{ii}\simeq
\frac{g_{ab}n_a}{L_z ( 1+2\alpha \sigma^2_{a} ) }
-\frac{\hbar\omega }{2}.
$$

\subsection{Integral $(II)$ (angular momentum)}
\label{sec:ang_mom_integral}

We then substitute the \textit{Ansatz} \eqref{eq:two_mode_expansion} in the integral $(II)$ and after  doing some derivations we find

\begin{align*}
    (II)&=-\Omega \int d^3 r\; \hat{\psi}_b^+ L_3 \hat{\psi}_b=\\
    &i\hbar\Omega\int d^3 r \sum_{i=1}^{2}\hat{b}_i^+ W_i(x\partial_y-y\partial_x)\sum_{j=1}^{2}\hat{b}_j W_j=\\
    &-2i\hbar\Omega L_z\sum_{i=1}^{2}\sum_{j=1}^{2}\frac{\gamma^2}{2}\hat{b}_i^+\hat{b}_j  e^{-\frac{\alpha}{2}(r^2_{v,i}+r^2_{v,j})+\alpha\;\bm r_{v,i}\cdot \bm r_{v,j} } \\
    &\bigg[ -\frac{x_{v,i}+x_{v,j}}{2}\pi\; y_{v,j}+\frac{ y_{v,i}+ y_{v,j}}{2}\pi \;x_{v,j} \bigg]
\end{align*}

For two co-rotating vortices such that $x_1=r_1\cos{\Theta}$,
$y_1=r_1\sin{\Theta}$, $x_2=-r_2\cos{\Theta}$, $y_2=-r_2\sin{\Theta}$, the factor in the square brackets is zero and $\mathcal{H}_0$ is rotation-invariant.

\begin{equation*}
   \bigg[ -\frac{x_{v,i}+x_{v,j}}{2}\pi\; y_{v,j}+\frac{ y_{v,i}+ y_{v,j}}{2}\pi \;x_{v,j} \bigg]=0
\end{equation*}

\subsection{Interaction term $\mathcal{U}$}

Let us compute the interaction term $\mathcal{U}$ by substituting the \textit{Ansatz} \eqref{eq:two_mode_expansion} in the relevant expression.

\begin{align*}
    \mathcal{U} &= \int d^3 r\; \frac{g_b}{2} (\hat{\psi}_b^+)^2\hat{\psi}_b^2 \\
    &= \frac{g_b}{2} \int d^3 r\;  \sum_{i=1}^{2} \hat{b}_i^+ W_i \sum_{j=1}^{2} \hat{b}_j^+ W_j \sum_{k=1}^{2} \hat{b}_k W_k \sum_{l=1}^{2} \hat{b}_l W_l \\
    & \simeq \frac{g_b}{2} \sum_{i=1}^{2} \int d^3 r\; W_i^4 \;(\hat{b}_i^+)^2 \hat{b}_i^2 =
    \frac{g_b}{2}  \sum_{i=1}^{2}K_{0 } \;\hat{n}_i(\hat{n}_i-1)
\end{align*}
where
$$K_{0 } = \int d^3 r\;W_i^4= \frac{\alpha }{\pi L_z} 
$$
Finally, the Hamiltonian $\mathcal{H}_0$ becomes, in the two-mode approximation

\begin{equation}
\begin{split}   
    \mathcal{H}_{tm} &= -\mu \sum_{i=1}^{2} \;\hat{n}_i - J\; (\hat{b}^+_{2}\hat{b}_1 -  \hat{b}^+_{1}\hat{b}_2)  + \frac{U}{2}  \sum_{i=1}^{2} \;\hat{n}_i(\hat{n}_i-1),
\end{split}    
\end{equation}

where $J=-|K_{12}|=-|K_{21}|$, $U=g_bK_0$, and $\mu=-E$.

\section{Mean-field approximation}
\label{sec:mean_field}

The procedure of Ref. \cite{FRANZOSI2000} for obtaining the mean-field Hamiltonian associated with $\mathcal{H}_{tm}$ assumes that bosons in the $i$-th
well are described by the
(local) coherent state $|z_i \rangle$, defined by
the equation $b_i |z_i \rangle= z_i |z_i \rangle$, related to  the annihilation operator $\hat{b}_i$ of the Weyl-Heisenberg algebra
$\{b_i, b_i^+, b^+_i b_i, {\mathbb I}_i \}$
at site $i$. The expectation values 
$$\langle z_i|\hat{b}_i|z_i\rangle=z_i, \quad 
\langle z_i|\hat{n}_i|z_i\rangle=|z_i|^2,
$$
show that $z_i$ can be seen as the order parameter describing the phase order at the $i$-th site, while 
$\langle z_i|\hat{n}_i|z_i\rangle$ describes the average boson population. The trial state of the variational coherent-state approach for the many-boson component $b$, can be written as
$$|\psi_b\rangle=e^{i \frac{S}{\hbar}}|Z\rangle, \quad
|Z\rangle=\otimes_{i=1,2}|z_i\rangle
$$
where
$$|z_i\rangle=e^{-\frac{|z_i|^2}{2}}\sum_{m=0}^{+\infty}\frac{z_i^m}{\sqrt{m!}}|m\rangle
,\quad z_i\in\mathbb{C}
$$
and $|m\rangle$ are the eigenstates of the number operator $b^+_i b_i$. By imposing the condition
$\langle \psi_b|( i\hbar \partial_\tau -\mathcal{H}_{tm} ) |\psi_b\rangle= 0$ one gets the 
the effective action

$$
S=\int_0^t d\tau\; L ,
\quad
L = i\hbar \langle Z|\partial_\tau|Z\rangle-\langle Z|\mathcal{H}_{tm}|Z\rangle,
$$
where $L$ represents the effective Lagrangian, and $H_{mf}=\langle Z|\mathcal{H}_{tm}|Z\rangle$, corresponds to the effective Hamiltonian.
For a symmetric BH dimer, we find
\begin{equation}
\begin{split}
    H_{mf}&=\langle Z|\mathcal{H}_{tm}|Z\rangle =\sum_{j=1}^2 \bigg(  \frac{U}{2}|z_j|^4-\mu|z_j|^2\bigg)+\\
    &-J\bigg(z_2^* z_1+z_1^* z_2\bigg)
\end{split}    
\end{equation}
The associated Euler-Lagrange equations are
\begin{align}
    &i\hbar\dot{z}_1=-\mu z_1+U|z_1|^2z_1-Jz_2\\
   &i\hbar\dot{z}_2=-\mu z_2+U|z_2|^2z_2-Jz_1,
\end{align}
while variables $z_1^*$, $z_2^*$ obey to the complex conjugate version of these equations. 
Interestingly, the same equations of motion can be derived by defining the  Poisson brackets 
$$
\left\{A,B \right\}=\frac{1}{i\hbar} \sum_{m=1}^2 \left[   \frac{\partial A}{\partial z_m}\frac{\partial B}{\partial z_m^*}- \frac{\partial A}{\partial z_m^*}\frac{\partial B}{\partial z_m}\right],
$$
showing how brackets $\{z_i, z^*_k\} = \delta_{ik}/(i\hbar)$ replace
commutators $\{b_i, b^+_k\} = \delta_{ik}$
in the derivation of the mean-field (semiclassical) picture of the BH dimer.
Not surprisingly, the particle number ${\cal N}= \langle Z|\sum_i b^+_ib_i|Z\rangle=
|z_1|^2+|z_2|^2$ represents a constant of motion, namely,
$\{ {\cal N}, {\cal H}_{mf}\}=0$.
Via an appropriate transformation, we find two new pairs of canonical variables replacing $z_1$, $z_1^*$, $z_2$, $z_2^*$. These are
$$
\mathcal{N}=|z_1|^2+|z_2|^2, \quad 
\psi=\frac{\phi_1+\phi_2}{2}
$$
$$
D=|z_1|^2-|z_2|^2, \quad
\theta=\frac{\phi_1-\phi_2}{2},
$$
where 
$z_j=|z_j|e^{i\phi_j}$
has been used.
Finally, the mean-field Hamiltonian in the new variable reads

\begin{equation}
    H_{mf}=\frac{U}{4}\mathcal{N}^2-\mu\mathcal{N}+\frac{U}{4}D^2-J\sqrt{\mathcal{N}^2-D^2}\cos{(2\theta)},
\end{equation}
associated with the Poisson Brackets

$$\left\{A,B\right\}=\frac{1}{\hbar}\left[ \frac{\partial A}{\partial D}\frac{\partial B}{\partial\theta}+\frac{\partial A}{\partial\mathcal{N}}\frac{\partial B}{\partial\psi} -\frac{\partial A}{\partial \theta}\frac{\partial B}{\partial D}-\frac{\partial A}{\partial\psi}\frac{\partial B}{\partial\mathcal{N}} \right],$$
satisfying  $\{\theta,\mathcal{N}\}=0=\{\psi,D\}$, $\{\theta, D\}=-\frac{1}{\hbar}=\{\psi,\mathcal{N}\}$. The relative equations of motion 
\begin{align}
    & \hbar\dot{D}=2J\sqrt{\mathcal{N}^2-D^2}\sin{(2\theta)}  \\
    & \hbar\dot{\theta}=-\frac{U}{2}D-\frac{JD}{\sqrt{\mathcal{N}^2-D^2}}\cos{(2\theta)} \\
    & \hbar\dot{\mathcal{N}}=0 
 \label{eq:Ndot}\\
    & \hbar\dot{\psi}=-\frac{U}
  {2}\mathcal{N}+\mu +\frac{J 
 \mathcal{N}}{\sqrt{\mathcal{N}^2-
 D^2}}\cos{(2\theta)} 
\label{eq:psidot}
\end{align}
feature the conserved quantity $\mathcal{N}$, corresponding to $N_b$, and show that the variable $\psi$ is auxiliary. Hence, the system is represented by the restricted phase space of the conjugate variables $(\theta, D)$, representing the phase shift of the two peaks and the population imbalance between the sites of the BJJ. Note that
the coherent state approach holds if the local boson population $\langle \hat{n}_i \rangle$ is large enough. The expectation values
\begin{align*}
    & \hat{n}_i \approx \langle  \hat{n}_i \rangle= \langle \hat{b}_i^+ \hat{b}_i \rangle = |z_i|^2\\
    & \hat{b}_i \approx \langle \hat{b}_i  \rangle = z_i  \\
    & \hat{b}_i^+ \approx \langle \hat{b}_i^+  \rangle = z_i^*    
\end{align*}
show how the coherent-state variational approach is equivalent to applying the Bogoliubov approximation.

\end{appendix}


\begin{thebibliography}{61}%
\makeatletter
\providecommand \@ifxundefined [1]{%
 \@ifx{#1\undefined}
}%
\providecommand \@ifnum [1]{%
 \ifnum #1\expandafter \@firstoftwo
 \else \expandafter \@secondoftwo
 \fi
}%
\providecommand \@ifx [1]{%
 \ifx #1\expandafter \@firstoftwo
 \else \expandafter \@secondoftwo
 \fi
}%
\providecommand \natexlab [1]{#1}%
\providecommand \enquote  [1]{``#1''}%
\providecommand \bibnamefont  [1]{#1}%
\providecommand \bibfnamefont [1]{#1}%
\providecommand \citenamefont [1]{#1}%
\providecommand \href@noop [0]{\@secondoftwo}%
\providecommand \href [0]{\begingroup \@sanitize@url \@href}%
\providecommand \@href[1]{\@@startlink{#1}\@@href}%
\providecommand \@@href[1]{\endgroup#1\@@endlink}%
\providecommand \@sanitize@url [0]{\catcode `\\12\catcode `\$12\catcode `\&12\catcode `\#12\catcode `\^12\catcode `\_12\catcode `\%12\relax}%
\providecommand \@@startlink[1]{}%
\providecommand \@@endlink[0]{}%
\providecommand \url  [0]{\begingroup\@sanitize@url \@url }%
\providecommand \@url [1]{\endgroup\@href {#1}{\urlprefix }}%
\providecommand \urlprefix  [0]{URL }%
\providecommand \Eprint [0]{\href }%
\providecommand \doibase [0]{https://doi.org/}%
\providecommand \selectlanguage [0]{\@gobble}%
\providecommand \bibinfo  [0]{\@secondoftwo}%
\providecommand \bibfield  [0]{\@secondoftwo}%
\providecommand \translation [1]{[#1]}%
\providecommand \BibitemOpen [0]{}%
\providecommand \bibitemStop [0]{}%
\providecommand \bibitemNoStop [0]{.\EOS\space}%
\providecommand \EOS [0]{\spacefactor3000\relax}%
\providecommand \BibitemShut  [1]{\csname bibitem#1\endcsname}%
\let\auto@bib@innerbib\@empty
\bibitem [{\citenamefont {Matthews}\ \emph {et~al.}(1999)\citenamefont {Matthews}, \citenamefont {Anderson}, \citenamefont {Haljan}, \citenamefont {Hall}, \citenamefont {Wieman},\ and\ \citenamefont {Cornell}}]{Matthews1999}%
  \BibitemOpen
  \bibfield  {author} {\bibinfo {author} {\bibfnamefont {M.~R.}\ \bibnamefont {Matthews}}, \bibinfo {author} {\bibfnamefont {B.~P.}\ \bibnamefont {Anderson}}, \bibinfo {author} {\bibfnamefont {P.~C.}\ \bibnamefont {Haljan}}, \bibinfo {author} {\bibfnamefont {D.~S.}\ \bibnamefont {Hall}}, \bibinfo {author} {\bibfnamefont {C.~E.}\ \bibnamefont {Wieman}},\ and\ \bibinfo {author} {\bibfnamefont {E.~A.}\ \bibnamefont {Cornell}},\ }\href {https://doi.org/10.1103/PhysRevLett.83.2498} {\bibfield  {journal} {\bibinfo  {journal} {Phys. Rev. Lett.}\ }\textbf {\bibinfo {volume} {83}},\ \bibinfo {pages} {2498} (\bibinfo {year} {1999})}\BibitemShut {NoStop}%
\bibitem [{\citenamefont {Anderson}\ \emph {et~al.}(2000)\citenamefont {Anderson}, \citenamefont {Haljan}, \citenamefont {Wieman},\ and\ \citenamefont {Cornell}}]{Anderson2000}%
  \BibitemOpen
  \bibfield  {author} {\bibinfo {author} {\bibfnamefont {B.~P.}\ \bibnamefont {Anderson}}, \bibinfo {author} {\bibfnamefont {P.~C.}\ \bibnamefont {Haljan}}, \bibinfo {author} {\bibfnamefont {C.~E.}\ \bibnamefont {Wieman}},\ and\ \bibinfo {author} {\bibfnamefont {E.~A.}\ \bibnamefont {Cornell}},\ }\href {https://doi.org/10.1103/PhysRevLett.85.2857} {\bibfield  {journal} {\bibinfo  {journal} {Phys. Rev. Lett.}\ }\textbf {\bibinfo {volume} {85}},\ \bibinfo {pages} {2857} (\bibinfo {year} {2000})}\BibitemShut {NoStop}%
\bibitem [{\citenamefont {Kuopanportti}\ \emph {et~al.}(2015)\citenamefont {Kuopanportti}, \citenamefont {Orlova},\ and\ \citenamefont {Milošević}}]{Kuopanportti2015}%
  \BibitemOpen
  \bibfield  {author} {\bibinfo {author} {\bibfnamefont {P.}~\bibnamefont {Kuopanportti}}, \bibinfo {author} {\bibfnamefont {N.~V.}\ \bibnamefont {Orlova}},\ and\ \bibinfo {author} {\bibfnamefont {M.~V.}\ \bibnamefont {Milošević}},\ }\href {https://doi.org/10.1103/PhysRevA.91.043605} {\bibfield  {journal} {\bibinfo  {journal} {Physical Review A}\ }\textbf {\bibinfo {volume} {91}},\ \bibinfo {pages} {043605} (\bibinfo {year} {2015})}\BibitemShut {NoStop}%
\bibitem [{\citenamefont {Patrick}\ \emph {et~al.}(2023)\citenamefont {Patrick}, \citenamefont {Gupta}, \citenamefont {Gregory},\ and\ \citenamefont {Barenghi}}]{Patrick2023}%
  \BibitemOpen
  \bibfield  {author} {\bibinfo {author} {\bibfnamefont {S.}~\bibnamefont {Patrick}}, \bibinfo {author} {\bibfnamefont {A.}~\bibnamefont {Gupta}}, \bibinfo {author} {\bibfnamefont {R.}~\bibnamefont {Gregory}},\ and\ \bibinfo {author} {\bibfnamefont {C.~F.}\ \bibnamefont {Barenghi}},\ }\href {https://doi.org/10.1103/PhysRevResearch.5.033201} {\bibfield  {journal} {\bibinfo  {journal} {Physical Review Research}\ }\textbf {\bibinfo {volume} {5}},\ \bibinfo {pages} {033201} (\bibinfo {year} {2023})}\BibitemShut {NoStop}%
\bibitem [{\citenamefont {Richaud}\ \emph {et~al.}(2023)\citenamefont {Richaud}, \citenamefont {Lamporesi}, \citenamefont {Capone},\ and\ \citenamefont {Recati}}]{Richaud2023}%
  \BibitemOpen
  \bibfield  {author} {\bibinfo {author} {\bibfnamefont {A.}~\bibnamefont {Richaud}}, \bibinfo {author} {\bibfnamefont {G.}~\bibnamefont {Lamporesi}}, \bibinfo {author} {\bibfnamefont {M.}~\bibnamefont {Capone}},\ and\ \bibinfo {author} {\bibfnamefont {A.}~\bibnamefont {Recati}},\ }\href {https://doi.org/10.1103/PhysRevA.107.053317} {\bibfield  {journal} {\bibinfo  {journal} {Physical Review A}\ }\textbf {\bibinfo {volume} {107}},\ \bibinfo {pages} {053317} (\bibinfo {year} {2023})}\BibitemShut {NoStop}%
\bibitem [{\citenamefont {Mueller}\ and\ \citenamefont {Ho}(2002)}]{Mueller2002}%
  \BibitemOpen
  \bibfield  {author} {\bibinfo {author} {\bibfnamefont {E.~J.}\ \bibnamefont {Mueller}}\ and\ \bibinfo {author} {\bibfnamefont {T.-L.}\ \bibnamefont {Ho}},\ }\href {https://doi.org/10.1103/PhysRevLett.88.180403} {\bibfield  {journal} {\bibinfo  {journal} {Physical Review Letters}\ }\textbf {\bibinfo {volume} {88}},\ \bibinfo {pages} {180403} (\bibinfo {year} {2002})}\BibitemShut {NoStop}%
\bibitem [{\citenamefont {Schweikhard}\ \emph {et~al.}(2004)\citenamefont {Schweikhard}, \citenamefont {Coddington}, \citenamefont {Engels}, \citenamefont {Tung}, ,\ and\ \citenamefont {Cornell}}]{Schweikhard2004}%
  \BibitemOpen
  \bibfield  {author} {\bibinfo {author} {\bibfnamefont {V.}~\bibnamefont {Schweikhard}}, \bibinfo {author} {\bibfnamefont {I.}~\bibnamefont {Coddington}}, \bibinfo {author} {\bibfnamefont {P.}~\bibnamefont {Engels}}, \bibinfo {author} {\bibfnamefont {S.}~\bibnamefont {Tung}}, ,\ and\ \bibinfo {author} {\bibfnamefont {E.~A.}\ \bibnamefont {Cornell}},\ }\href {https://doi.org/10.1103/PhysRevLett.93.210403} {\bibfield  {journal} {\bibinfo  {journal} {Physical Review Letters}\ }\textbf {\bibinfo {volume} {93}},\ \bibinfo {pages} {210403} (\bibinfo {year} {2004})}\BibitemShut {NoStop}%
\bibitem [{\citenamefont {Kuopanportti}\ \emph {et~al.}(2012)\citenamefont {Kuopanportti}, \citenamefont {Huhtamäki},\ and\ \citenamefont {Möttönen}}]{Kuopanportti2012}%
  \BibitemOpen
  \bibfield  {author} {\bibinfo {author} {\bibfnamefont {P.}~\bibnamefont {Kuopanportti}}, \bibinfo {author} {\bibfnamefont {J.~A.~M.}\ \bibnamefont {Huhtamäki}},\ and\ \bibinfo {author} {\bibfnamefont {M.}~\bibnamefont {Möttönen}},\ }\href {https://doi.org/10.1103/PhysRevA.85.043613} {\bibfield  {journal} {\bibinfo  {journal} {Physical Review A}\ }\textbf {\bibinfo {volume} {85}},\ \bibinfo {pages} {043613} (\bibinfo {year} {2012})}\BibitemShut {NoStop}%
\bibitem [{\citenamefont {Kuopanportti}\ \emph {et~al.}(2019)\citenamefont {Kuopanportti}, \citenamefont {Bandyopadhyay}, \citenamefont {Roy},\ and\ \citenamefont {Angom}}]{Kuopanportti2019}%
  \BibitemOpen
  \bibfield  {author} {\bibinfo {author} {\bibfnamefont {P.}~\bibnamefont {Kuopanportti}}, \bibinfo {author} {\bibfnamefont {S.}~\bibnamefont {Bandyopadhyay}}, \bibinfo {author} {\bibfnamefont {A.}~\bibnamefont {Roy}},\ and\ \bibinfo {author} {\bibfnamefont {D.}~\bibnamefont {Angom}},\ }\href {https://doi.org/10.1103/PhysRevA.100.033615} {\bibfield  {journal} {\bibinfo  {journal} {Phys. Rev. A}\ }\textbf {\bibinfo {volume} {100}},\ \bibinfo {pages} {033615} (\bibinfo {year} {2019})}\BibitemShut {NoStop}%
\bibitem [{\citenamefont {Richaud}\ \emph {et~al.}(2020)\citenamefont {Richaud}, \citenamefont {Penna}, \citenamefont {Mayol},\ and\ \citenamefont {Guilleumas}}]{Richaud2020}%
  \BibitemOpen
  \bibfield  {author} {\bibinfo {author} {\bibfnamefont {A.}~\bibnamefont {Richaud}}, \bibinfo {author} {\bibfnamefont {V.}~\bibnamefont {Penna}}, \bibinfo {author} {\bibfnamefont {R.}~\bibnamefont {Mayol}},\ and\ \bibinfo {author} {\bibfnamefont {M.}~\bibnamefont {Guilleumas}},\ }\href {https://doi.org/10.1103/PhysRevA.101.013630} {\bibfield  {journal} {\bibinfo  {journal} {Physical Review A}\ }\textbf {\bibinfo {volume} {101}},\ \bibinfo {pages} {013630} (\bibinfo {year} {2020})}\BibitemShut {NoStop}%
\bibitem [{\citenamefont {Bellettini}\ \emph {et~al.}(2023)\citenamefont {Bellettini}, \citenamefont {Richaud},\ and\ \citenamefont {Penna}}]{Bellettini2023}%
  \BibitemOpen
  \bibfield  {author} {\bibinfo {author} {\bibfnamefont {A.}~\bibnamefont {Bellettini}}, \bibinfo {author} {\bibfnamefont {A.}~\bibnamefont {Richaud}},\ and\ \bibinfo {author} {\bibfnamefont {V.}~\bibnamefont {Penna}},\ }\href {https://doi.org/10.1140/epjp/s13360-023-04294-6} {\bibfield  {journal} {\bibinfo  {journal} {The European Physical Journal Plus}\ }\textbf {\bibinfo {volume} {138}},\ \bibinfo {pages} {676} (\bibinfo {year} {2023})}\BibitemShut {NoStop}%
\bibitem [{\citenamefont {Bellettini}\ \emph {et~al.}(2024)\citenamefont {Bellettini}, \citenamefont {Richaud},\ and\ \citenamefont {Penna}}]{Bellettini2024}%
  \BibitemOpen
  \bibfield  {author} {\bibinfo {author} {\bibfnamefont {A.}~\bibnamefont {Bellettini}}, \bibinfo {author} {\bibfnamefont {A.}~\bibnamefont {Richaud}},\ and\ \bibinfo {author} {\bibfnamefont {V.}~\bibnamefont {Penna}},\ }\href {https://doi.org/10.1103/PhysRevA.109.053301} {\bibfield  {journal} {\bibinfo  {journal} {Phys. Rev. A}\ }\textbf {\bibinfo {volume} {109}},\ \bibinfo {pages} {053301} (\bibinfo {year} {2024})}\BibitemShut {NoStop}%
\bibitem [{\citenamefont {White}\ \emph {et~al.}(2006)\citenamefont {White}, \citenamefont {Gao}, \citenamefont {Pasienski},\ and\ \citenamefont {DeMarco}}]{White2006}%
  \BibitemOpen
  \bibfield  {author} {\bibinfo {author} {\bibfnamefont {M.}~\bibnamefont {White}}, \bibinfo {author} {\bibfnamefont {H.}~\bibnamefont {Gao}}, \bibinfo {author} {\bibfnamefont {M.}~\bibnamefont {Pasienski}},\ and\ \bibinfo {author} {\bibfnamefont {B.}~\bibnamefont {DeMarco}},\ }\href {https://doi.org/10.1103/PhysRevA.74.023616} {\bibfield  {journal} {\bibinfo  {journal} {Phys. Rev. A}\ }\textbf {\bibinfo {volume} {74}},\ \bibinfo {pages} {023616} (\bibinfo {year} {2006})}\BibitemShut {NoStop}%
\bibitem [{\citenamefont {Navon}\ \emph {et~al.}(2021)\citenamefont {Navon}, \citenamefont {Smith},\ and\ \citenamefont {Hadzibabic}}]{Navon2021}%
  \BibitemOpen
  \bibfield  {author} {\bibinfo {author} {\bibfnamefont {N.}~\bibnamefont {Navon}}, \bibinfo {author} {\bibfnamefont {R.~P.}\ \bibnamefont {Smith}},\ and\ \bibinfo {author} {\bibfnamefont {Z.}~\bibnamefont {Hadzibabic}},\ }\href {https://doi.org/10.1038/s41567-021-01403-z} {\bibfield  {journal} {\bibinfo  {journal} {Nature Physics}\ }\textbf {\bibinfo {volume} {17}},\ \bibinfo {pages} {1334} (\bibinfo {year} {2021})}\BibitemShut {NoStop}%
\bibitem [{\citenamefont {Caldara}\ \emph {et~al.}(2023)\citenamefont {Caldara}, \citenamefont {Richaud}, \citenamefont {Capone},\ and\ \citenamefont {Massignan}}]{Caldara2023}%
  \BibitemOpen
  \bibfield  {author} {\bibinfo {author} {\bibfnamefont {M.}~\bibnamefont {Caldara}}, \bibinfo {author} {\bibfnamefont {A.}~\bibnamefont {Richaud}}, \bibinfo {author} {\bibfnamefont {M.}~\bibnamefont {Capone}},\ and\ \bibinfo {author} {\bibfnamefont {P.}~\bibnamefont {Massignan}},\ }\href {https://doi.org/10.21468/SciPostPhys.15.2.057} {\bibfield  {journal} {\bibinfo  {journal} {SciPost Physics}\ }\textbf {\bibinfo {volume} {15}},\ \bibinfo {pages} {057} (\bibinfo {year} {2023})}\BibitemShut {NoStop}%
\bibitem [{\citenamefont {Lundblad}\ \emph {et~al.}(2019)\citenamefont {Lundblad}, \citenamefont {Carollo}, \citenamefont {Lannert}, \citenamefont {Gold}, \citenamefont {Jiang}, \citenamefont {Paseltiner}, \citenamefont {Sergay},\ and\ \citenamefont {Aveline}}]{Lundblad2019}%
  \BibitemOpen
  \bibfield  {author} {\bibinfo {author} {\bibfnamefont {N.}~\bibnamefont {Lundblad}}, \bibinfo {author} {\bibfnamefont {R.~A.}\ \bibnamefont {Carollo}}, \bibinfo {author} {\bibfnamefont {C.}~\bibnamefont {Lannert}}, \bibinfo {author} {\bibfnamefont {M.~J.}\ \bibnamefont {Gold}}, \bibinfo {author} {\bibfnamefont {X.}~\bibnamefont {Jiang}}, \bibinfo {author} {\bibfnamefont {D.}~\bibnamefont {Paseltiner}}, \bibinfo {author} {\bibfnamefont {N.}~\bibnamefont {Sergay}},\ and\ \bibinfo {author} {\bibfnamefont {D.~C.}\ \bibnamefont {Aveline}},\ }\href {https://doi.org/10.1038/s41526-019-0087-y} {\bibfield  {journal} {\bibinfo  {journal} {npj Microgravity}\ }\textbf {\bibinfo {volume} {5}},\ \bibinfo {pages} {30} (\bibinfo {year} {2019})}\BibitemShut {NoStop}%
\bibitem [{\citenamefont {Guo}\ \emph {et~al.}(2022)\citenamefont {Guo}, \citenamefont {Gutierrez}, \citenamefont {Rey}, \citenamefont {Badr}, \citenamefont {Perrin}, \citenamefont {Longchambon}, \citenamefont {Bagnato}, \citenamefont {Perrin},\ and\ \citenamefont {Dubessy}}]{Guo2022}%
  \BibitemOpen
  \bibfield  {author} {\bibinfo {author} {\bibfnamefont {Y.}~\bibnamefont {Guo}}, \bibinfo {author} {\bibfnamefont {E.~M.}\ \bibnamefont {Gutierrez}}, \bibinfo {author} {\bibfnamefont {D.}~\bibnamefont {Rey}}, \bibinfo {author} {\bibfnamefont {T.}~\bibnamefont {Badr}}, \bibinfo {author} {\bibfnamefont {A.}~\bibnamefont {Perrin}}, \bibinfo {author} {\bibfnamefont {L.}~\bibnamefont {Longchambon}}, \bibinfo {author} {\bibfnamefont {V.~S.}\ \bibnamefont {Bagnato}}, \bibinfo {author} {\bibfnamefont {H.}~\bibnamefont {Perrin}},\ and\ \bibinfo {author} {\bibfnamefont {R.}~\bibnamefont {Dubessy}},\ }\href {https://doi.org/10.1088/1367-2630/ac919f} {\bibfield  {journal} {\bibinfo  {journal} {New Journal of Physics}\ }\textbf {\bibinfo {volume} {24}},\ \bibinfo {pages} {093040} (\bibinfo {year} {2022})}\BibitemShut {NoStop}%
\bibitem [{\citenamefont {Wolf}\ \emph {et~al.}(2022)\citenamefont {Wolf}, \citenamefont {Boegel}, \citenamefont {Meister}, \citenamefont {Balaž}, \citenamefont {Gaaloul},\ and\ \citenamefont {Efremov}}]{Wolf2022}%
  \BibitemOpen
  \bibfield  {author} {\bibinfo {author} {\bibfnamefont {A.}~\bibnamefont {Wolf}}, \bibinfo {author} {\bibfnamefont {P.}~\bibnamefont {Boegel}}, \bibinfo {author} {\bibfnamefont {M.}~\bibnamefont {Meister}}, \bibinfo {author} {\bibfnamefont {A.}~\bibnamefont {Balaž}}, \bibinfo {author} {\bibfnamefont {N.}~\bibnamefont {Gaaloul}},\ and\ \bibinfo {author} {\bibfnamefont {M.~A.}\ \bibnamefont {Efremov}},\ }\href {https://doi.org/10.1103/PhysRevA.106.013309} {\bibfield  {journal} {\bibinfo  {journal} {Physical Review A}\ }\textbf {\bibinfo {volume} {106}},\ \bibinfo {pages} {013309} (\bibinfo {year} {2022})}\BibitemShut {NoStop}%
\bibitem [{\citenamefont {Móller}\ \emph {et~al.}(2020)\citenamefont {Móller}, \citenamefont {dos Santos}, \citenamefont {Bagnato},\ and\ \citenamefont {Pelster}}]{Móller2020}%
  \BibitemOpen
  \bibfield  {author} {\bibinfo {author} {\bibfnamefont {N.~S.}\ \bibnamefont {Móller}}, \bibinfo {author} {\bibfnamefont {F.~E.~A.}\ \bibnamefont {dos Santos}}, \bibinfo {author} {\bibfnamefont {V.~S.}\ \bibnamefont {Bagnato}},\ and\ \bibinfo {author} {\bibfnamefont {A.}~\bibnamefont {Pelster}},\ }\href {https://doi.org/10.1088/1367-2630/ab91fb} {\bibfield  {journal} {\bibinfo  {journal} {New Journal of Physics}\ }\textbf {\bibinfo {volume} {22}},\ \bibinfo {pages} {063059} (\bibinfo {year} {2020})}\BibitemShut {NoStop}%
\bibitem [{\citenamefont {Caracanhas}\ \emph {et~al.}(2022)\citenamefont {Caracanhas}, \citenamefont {Massignan},\ and\ \citenamefont {Fetter}}]{Caracanhas2022}%
  \BibitemOpen
  \bibfield  {author} {\bibinfo {author} {\bibfnamefont {M.~A.}\ \bibnamefont {Caracanhas}}, \bibinfo {author} {\bibfnamefont {P.}~\bibnamefont {Massignan}},\ and\ \bibinfo {author} {\bibfnamefont {A.~L.}\ \bibnamefont {Fetter}},\ }\href {https://doi.org/10.1103/PhysRevA.105.023307} {\bibfield  {journal} {\bibinfo  {journal} {Phys. Rev. A}\ }\textbf {\bibinfo {volume} {105}},\ \bibinfo {pages} {023307} (\bibinfo {year} {2022})}\BibitemShut {NoStop}%
\bibitem [{\citenamefont {Davis}\ \emph {et~al.}(2009)\citenamefont {Davis}, \citenamefont {Carretero-Gonz\'alez}, \citenamefont {Shi}, \citenamefont {Law}, \citenamefont {Kevrekidis},\ and\ \citenamefont {Anderson}}]{Davis2009}%
  \BibitemOpen
  \bibfield  {author} {\bibinfo {author} {\bibfnamefont {M.~C.}\ \bibnamefont {Davis}}, \bibinfo {author} {\bibfnamefont {R.}~\bibnamefont {Carretero-Gonz\'alez}}, \bibinfo {author} {\bibfnamefont {Z.}~\bibnamefont {Shi}}, \bibinfo {author} {\bibfnamefont {K.~J.~H.}\ \bibnamefont {Law}}, \bibinfo {author} {\bibfnamefont {P.~G.}\ \bibnamefont {Kevrekidis}},\ and\ \bibinfo {author} {\bibfnamefont {B.~P.}\ \bibnamefont {Anderson}},\ }\href {https://doi.org/10.1103/PhysRevA.80.023604} {\bibfield  {journal} {\bibinfo  {journal} {Phys. Rev. A}\ }\textbf {\bibinfo {volume} {80}},\ \bibinfo {pages} {023604} (\bibinfo {year} {2009})}\BibitemShut {NoStop}%
\bibitem [{\citenamefont {Samson}\ \emph {et~al.}(2016)\citenamefont {Samson}, \citenamefont {Wilson}, \citenamefont {Newman},\ and\ \citenamefont {Anderson}}]{Samson2016}%
  \BibitemOpen
  \bibfield  {author} {\bibinfo {author} {\bibfnamefont {E.~C.}\ \bibnamefont {Samson}}, \bibinfo {author} {\bibfnamefont {K.~E.}\ \bibnamefont {Wilson}}, \bibinfo {author} {\bibfnamefont {Z.~L.}\ \bibnamefont {Newman}},\ and\ \bibinfo {author} {\bibfnamefont {B.~P.}\ \bibnamefont {Anderson}},\ }\href {https://doi.org/10.1103/PhysRevA.93.023603} {\bibfield  {journal} {\bibinfo  {journal} {Phys. Rev. A}\ }\textbf {\bibinfo {volume} {93}},\ \bibinfo {pages} {023603} (\bibinfo {year} {2016})}\BibitemShut {NoStop}%
\bibitem [{\citenamefont {Gallemí}\ \emph {et~al.}(2018)\citenamefont {Gallemí}, \citenamefont {Pitaevskii}, \citenamefont {Stringari},\ and\ \citenamefont {Recati}}]{Gallemí2018}%
  \BibitemOpen
  \bibfield  {author} {\bibinfo {author} {\bibfnamefont {A.}~\bibnamefont {Gallemí}}, \bibinfo {author} {\bibfnamefont {L.~P.}\ \bibnamefont {Pitaevskii}}, \bibinfo {author} {\bibfnamefont {S.}~\bibnamefont {Stringari}},\ and\ \bibinfo {author} {\bibfnamefont {A.}~\bibnamefont {Recati}},\ }\href {https://doi.org/10.1103/PhysRevA.97.063615} {\bibfield  {journal} {\bibinfo  {journal} {Physical Review A}\ }\textbf {\bibinfo {volume} {97}},\ \bibinfo {pages} {063615} (\bibinfo {year} {2018})}\BibitemShut {NoStop}%
\bibitem [{\citenamefont {Onsager}(1949)}]{Onsager1949}%
  \BibitemOpen
  \bibfield  {author} {\bibinfo {author} {\bibfnamefont {L.}~\bibnamefont {Onsager}},\ }\href {https://doi.org/10.1007/BF02780991} {\bibfield  {journal} {\bibinfo  {journal} {Il Nuovo Cimento}\ }\textbf {\bibinfo {volume} {6}},\ \bibinfo {pages} {279} (\bibinfo {year} {1949})}\BibitemShut {NoStop}%
\bibitem [{\citenamefont {Fiszdon}(1991)}]{Fiszdon1991}%
  \BibitemOpen
  \bibfield  {author} {\bibinfo {author} {\bibfnamefont {W.}~\bibnamefont {Fiszdon}},\ }\href {https://doi.org/10.1017/S0022112091220650} {\bibfield  {journal} {\bibinfo  {journal} {Journal of Fluid Mechanics}\ }\textbf {\bibinfo {volume} {233}},\ \bibinfo {pages} {691} (\bibinfo {year} {1991})}\BibitemShut {NoStop}%
\bibitem [{\citenamefont {McGee}\ and\ \citenamefont {Holland}(2001)}]{McGee2001}%
  \BibitemOpen
  \bibfield  {author} {\bibinfo {author} {\bibfnamefont {S.~A.}\ \bibnamefont {McGee}}\ and\ \bibinfo {author} {\bibfnamefont {M.~J.}\ \bibnamefont {Holland}},\ }\href {https://doi.org/10.1103/PhysRevA.63.043608} {\bibfield  {journal} {\bibinfo  {journal} {Phys. Rev. A}\ }\textbf {\bibinfo {volume} {63}},\ \bibinfo {pages} {043608} (\bibinfo {year} {2001})}\BibitemShut {NoStop}%
\bibitem [{\citenamefont {Law}\ \emph {et~al.}(2010)\citenamefont {Law}, \citenamefont {Kevrekidis},\ and\ \citenamefont {Tuckerman}}]{Law2010}%
  \BibitemOpen
  \bibfield  {author} {\bibinfo {author} {\bibfnamefont {K.~J.~H.}\ \bibnamefont {Law}}, \bibinfo {author} {\bibfnamefont {P.~G.}\ \bibnamefont {Kevrekidis}},\ and\ \bibinfo {author} {\bibfnamefont {L.~S.}\ \bibnamefont {Tuckerman}},\ }\href {https://doi.org/10.1103/PhysRevLett.105.160405} {\bibfield  {journal} {\bibinfo  {journal} {Phys. Rev. Lett.}\ }\textbf {\bibinfo {volume} {105}},\ \bibinfo {pages} {160405} (\bibinfo {year} {2010})}\BibitemShut {NoStop}%
\bibitem [{\citenamefont {Richaud}\ \emph {et~al.}(2021)\citenamefont {Richaud}, \citenamefont {Penna},\ and\ \citenamefont {Fetter}}]{Richaud2021}%
  \BibitemOpen
  \bibfield  {author} {\bibinfo {author} {\bibfnamefont {A.}~\bibnamefont {Richaud}}, \bibinfo {author} {\bibfnamefont {V.}~\bibnamefont {Penna}},\ and\ \bibinfo {author} {\bibfnamefont {A.~L.}\ \bibnamefont {Fetter}},\ }\href {https://doi.org/10.1103/PhysRevA.103.023311} {\bibfield  {journal} {\bibinfo  {journal} {Physical Review A}\ }\textbf {\bibinfo {volume} {103}},\ \bibinfo {pages} {023311} (\bibinfo {year} {2021})}\BibitemShut {NoStop}%
\bibitem [{\citenamefont {Milburn}\ \emph {et~al.}(1997)\citenamefont {Milburn}, \citenamefont {Corney}, \citenamefont {Wright},\ and\ \citenamefont {Walls}}]{Milburn1997}%
  \BibitemOpen
  \bibfield  {author} {\bibinfo {author} {\bibfnamefont {G.~J.}\ \bibnamefont {Milburn}}, \bibinfo {author} {\bibfnamefont {J.}~\bibnamefont {Corney}}, \bibinfo {author} {\bibfnamefont {E.~M.}\ \bibnamefont {Wright}},\ and\ \bibinfo {author} {\bibfnamefont {D.~F.}\ \bibnamefont {Walls}},\ }\href {https://doi.org/10.1103/PhysRevA.55.4318} {\bibfield  {journal} {\bibinfo  {journal} {Phys. Rev. A}\ }\textbf {\bibinfo {volume} {55}},\ \bibinfo {pages} {4318} (\bibinfo {year} {1997})}\BibitemShut {NoStop}%
\bibitem [{\citenamefont {Smerzi}\ \emph {et~al.}(1997)\citenamefont {Smerzi}, \citenamefont {Fantoni}, \citenamefont {Giovanazzi},\ and\ \citenamefont {Shenoy}}]{Smerzi1997}%
  \BibitemOpen
  \bibfield  {author} {\bibinfo {author} {\bibfnamefont {A.}~\bibnamefont {Smerzi}}, \bibinfo {author} {\bibfnamefont {S.}~\bibnamefont {Fantoni}}, \bibinfo {author} {\bibfnamefont {S.}~\bibnamefont {Giovanazzi}},\ and\ \bibinfo {author} {\bibfnamefont {S.~R.}\ \bibnamefont {Shenoy}},\ }\href {https://doi.org/10.1103/PhysRevLett.79.4950} {\bibfield  {journal} {\bibinfo  {journal} {Phys. Rev. Lett.}\ }\textbf {\bibinfo {volume} {79}},\ \bibinfo {pages} {4950} (\bibinfo {year} {1997})}\BibitemShut {NoStop}%
\bibitem [{\citenamefont {Raghavan}\ \emph {et~al.}(1999)\citenamefont {Raghavan}, \citenamefont {Smerzi}, \citenamefont {Fantoni},\ and\ \citenamefont {Shenoy}}]{Raghavan1999}%
  \BibitemOpen
  \bibfield  {author} {\bibinfo {author} {\bibfnamefont {S.}~\bibnamefont {Raghavan}}, \bibinfo {author} {\bibfnamefont {A.}~\bibnamefont {Smerzi}}, \bibinfo {author} {\bibfnamefont {S.}~\bibnamefont {Fantoni}},\ and\ \bibinfo {author} {\bibfnamefont {S.~R.}\ \bibnamefont {Shenoy}},\ }\href {https://doi.org/10.1103/PhysRevA.59.620} {\bibfield  {journal} {\bibinfo  {journal} {Phys. Rev. A}\ }\textbf {\bibinfo {volume} {59}},\ \bibinfo {pages} {620} (\bibinfo {year} {1999})}\BibitemShut {NoStop}%
\bibitem [{\citenamefont {Giovanazzi}\ \emph {et~al.}(2000)\citenamefont {Giovanazzi}, \citenamefont {Smerzi},\ and\ \citenamefont {Fantoni}}]{Giovanazzi2000}%
  \BibitemOpen
  \bibfield  {author} {\bibinfo {author} {\bibfnamefont {S.}~\bibnamefont {Giovanazzi}}, \bibinfo {author} {\bibfnamefont {A.}~\bibnamefont {Smerzi}},\ and\ \bibinfo {author} {\bibfnamefont {S.}~\bibnamefont {Fantoni}},\ }\href {https://doi.org/10.1103/PhysRevLett.84.4521} {\bibfield  {journal} {\bibinfo  {journal} {Phys. Rev. Lett.}\ }\textbf {\bibinfo {volume} {84}},\ \bibinfo {pages} {4521} (\bibinfo {year} {2000})}\BibitemShut {NoStop}%
\bibitem [{\citenamefont {Franzosi}\ \emph {et~al.}(2000)\citenamefont {Franzosi}, \citenamefont {Penna},\ and\ \citenamefont {Zecchina}}]{FRANZOSI2000}%
  \BibitemOpen
  \bibfield  {author} {\bibinfo {author} {\bibfnamefont {R.}~\bibnamefont {Franzosi}}, \bibinfo {author} {\bibfnamefont {V.}~\bibnamefont {Penna}},\ and\ \bibinfo {author} {\bibfnamefont {R.}~\bibnamefont {Zecchina}},\ }\href {https://doi.org/10.1142/S0217979200001011} {\bibfield  {journal} {\bibinfo  {journal} {International Journal of Modern Physics B}\ }\textbf {\bibinfo {volume} {14}},\ \bibinfo {pages} {943} (\bibinfo {year} {2000})}\BibitemShut {NoStop}%
\bibitem [{\citenamefont {Ananikian}\ and\ \citenamefont {Bergeman}(2006{\natexlab{a}})}]{ana2006}%
  \BibitemOpen
  \bibfield  {author} {\bibinfo {author} {\bibfnamefont {D.}~\bibnamefont {Ananikian}}\ and\ \bibinfo {author} {\bibfnamefont {T.}~\bibnamefont {Bergeman}},\ }\href {https://doi.org/10.1103/PhysRevA.73.013604} {\bibfield  {journal} {\bibinfo  {journal} {Physical Review A}\ }\textbf {\bibinfo {volume} {73}},\ \bibinfo {pages} {013604} (\bibinfo {year} {2006}{\natexlab{a}})}\BibitemShut {NoStop}%
\bibitem [{\citenamefont {Ferrini}\ \emph {et~al.}(2008)\citenamefont {Ferrini}, \citenamefont {Minguzzi},\ and\ \citenamefont {Hekking}}]{ming2008}%
  \BibitemOpen
  \bibfield  {author} {\bibinfo {author} {\bibfnamefont {G.}~\bibnamefont {Ferrini}}, \bibinfo {author} {\bibfnamefont {A.}~\bibnamefont {Minguzzi}},\ and\ \bibinfo {author} {\bibfnamefont {F.~W.~J.}\ \bibnamefont {Hekking}},\ }\href {https://doi.org/10.1103/PhysRevA.78.023606} {\bibfield  {journal} {\bibinfo  {journal} {Physical Review A}\ }\textbf {\bibinfo {volume} {78}},\ \bibinfo {pages} {023606} (\bibinfo {year} {2008})}\BibitemShut {NoStop}%
\bibitem [{\citenamefont {Sakmann}\ \emph {et~al.}(2009)\citenamefont {Sakmann}, \citenamefont {Streltsov}, \citenamefont {Alon},\ and\ \citenamefont {Cederbaum}}]{alon2009}%
  \BibitemOpen
  \bibfield  {author} {\bibinfo {author} {\bibfnamefont {K.}~\bibnamefont {Sakmann}}, \bibinfo {author} {\bibfnamefont {A.~I.}\ \bibnamefont {Streltsov}}, \bibinfo {author} {\bibfnamefont {O.~E.}\ \bibnamefont {Alon}},\ and\ \bibinfo {author} {\bibfnamefont {L.~S.}\ \bibnamefont {Cederbaum}},\ }\href {https://doi.org/10.1103/PhysRevLett.103.220601} {\bibfield  {journal} {\bibinfo  {journal} {Phys. Rev. Lett.}\ }\textbf {\bibinfo {volume} {103}},\ \bibinfo {pages} {220601} (\bibinfo {year} {2009})}\BibitemShut {NoStop}%
\bibitem [{\citenamefont {Chuchem}\ \emph {et~al.}(2010)\citenamefont {Chuchem}, \citenamefont {Smith-Mannschott}, \citenamefont {Hiller}, \citenamefont {Kottos}, \citenamefont {Vardi},\ and\ \citenamefont {Cohen}}]{Chuchem2010}%
  \BibitemOpen
  \bibfield  {author} {\bibinfo {author} {\bibfnamefont {M.}~\bibnamefont {Chuchem}}, \bibinfo {author} {\bibfnamefont {K.}~\bibnamefont {Smith-Mannschott}}, \bibinfo {author} {\bibfnamefont {M.}~\bibnamefont {Hiller}}, \bibinfo {author} {\bibfnamefont {T.}~\bibnamefont {Kottos}}, \bibinfo {author} {\bibfnamefont {A.}~\bibnamefont {Vardi}},\ and\ \bibinfo {author} {\bibfnamefont {D.}~\bibnamefont {Cohen}},\ }\href {https://doi.org/10.1103/PhysRevA.82.053617} {\bibfield  {journal} {\bibinfo  {journal} {Physical Review A}\ }\textbf {\bibinfo {volume} {82}},\ \bibinfo {pages} {053617} (\bibinfo {year} {2010})}\BibitemShut {NoStop}%
\bibitem [{\citenamefont {Xhani}\ \emph {et~al.}(2020)\citenamefont {Xhani}, \citenamefont {Galantucci}, \citenamefont {Barenghi}, \citenamefont {Roati}, \citenamefont {Trombettoni},\ and\ \citenamefont {Proukakis}}]{Xhani2020}%
  \BibitemOpen
  \bibfield  {author} {\bibinfo {author} {\bibfnamefont {K.}~\bibnamefont {Xhani}}, \bibinfo {author} {\bibfnamefont {L.}~\bibnamefont {Galantucci}}, \bibinfo {author} {\bibfnamefont {C.~F.}\ \bibnamefont {Barenghi}}, \bibinfo {author} {\bibfnamefont {G.}~\bibnamefont {Roati}}, \bibinfo {author} {\bibfnamefont {A.}~\bibnamefont {Trombettoni}},\ and\ \bibinfo {author} {\bibfnamefont {N.~P.}\ \bibnamefont {Proukakis}},\ }\href {https://doi.org/10.1088/1367-2630/abc8e4} {\bibfield  {journal} {\bibinfo  {journal} {New Journal of Physics}\ }\textbf {\bibinfo {volume} {22}},\ \bibinfo {pages} {123006} (\bibinfo {year} {2020})}\BibitemShut {NoStop}%
\bibitem [{\citenamefont {Amico}\ \emph {et~al.}(2017)\citenamefont {Amico}, \citenamefont {Birkl}, \citenamefont {Boshier},\ and\ \citenamefont {Kwek}}]{Amico2017}%
  \BibitemOpen
  \bibfield  {author} {\bibinfo {author} {\bibfnamefont {L.}~\bibnamefont {Amico}}, \bibinfo {author} {\bibfnamefont {G.}~\bibnamefont {Birkl}}, \bibinfo {author} {\bibfnamefont {M.}~\bibnamefont {Boshier}},\ and\ \bibinfo {author} {\bibfnamefont {L.-C.}\ \bibnamefont {Kwek}},\ }\href {https://doi.org/10.1088/1367-2630/aa5a6d} {\bibfield  {journal} {\bibinfo  {journal} {New Journal of Physics}\ }\textbf {\bibinfo {volume} {19}},\ \bibinfo {pages} {020201} (\bibinfo {year} {2017})}\BibitemShut {NoStop}%
\bibitem [{\citenamefont {Mazzarella}\ \emph {et~al.}(2009)\citenamefont {Mazzarella}, \citenamefont {Moratti}, \citenamefont {Salasnich}, \citenamefont {Salerno},\ and\ \citenamefont {Toigo}}]{Mazzarella2009}%
  \BibitemOpen
  \bibfield  {author} {\bibinfo {author} {\bibfnamefont {G.}~\bibnamefont {Mazzarella}}, \bibinfo {author} {\bibfnamefont {M.}~\bibnamefont {Moratti}}, \bibinfo {author} {\bibfnamefont {L.}~\bibnamefont {Salasnich}}, \bibinfo {author} {\bibfnamefont {M.}~\bibnamefont {Salerno}},\ and\ \bibinfo {author} {\bibfnamefont {F.}~\bibnamefont {Toigo}},\ }\href {https://doi.org/10.1088/0953-4075/42/12/125301} {\bibfield  {journal} {\bibinfo  {journal} {Journal of Physics B: Atomic, Molecular and Optical Physics}\ }\textbf {\bibinfo {volume} {42}},\ \bibinfo {pages} {125301} (\bibinfo {year} {2009})}\BibitemShut {NoStop}%
\bibitem [{\citenamefont {Lingua}\ \emph {et~al.}(2016)\citenamefont {Lingua}, \citenamefont {Mazzarella},\ and\ \citenamefont {Penna}}]{Lingua2016}%
  \BibitemOpen
  \bibfield  {author} {\bibinfo {author} {\bibfnamefont {F.}~\bibnamefont {Lingua}}, \bibinfo {author} {\bibfnamefont {G.}~\bibnamefont {Mazzarella}},\ and\ \bibinfo {author} {\bibfnamefont {V.}~\bibnamefont {Penna}},\ }\href {https://doi.org/10.1088/0953-4075/49/20/205005} {\bibfield  {journal} {\bibinfo  {journal} {Journal of Physics B: Atomic, Molecular and Optical Physics}\ }\textbf {\bibinfo {volume} {49}},\ \bibinfo {pages} {205005} (\bibinfo {year} {2016})}\BibitemShut {NoStop}%
\bibitem [{\citenamefont {Penna}\ and\ \citenamefont {Richaud}(2017)}]{Penna2017}%
  \BibitemOpen
  \bibfield  {author} {\bibinfo {author} {\bibfnamefont {V.}~\bibnamefont {Penna}}\ and\ \bibinfo {author} {\bibfnamefont {A.}~\bibnamefont {Richaud}},\ }\href {https://doi.org/10.1103/PhysRevA.96.053631} {\bibfield  {journal} {\bibinfo  {journal} {Physical Review A}\ }\textbf {\bibinfo {volume} {96}},\ \bibinfo {pages} {053631} (\bibinfo {year} {2017})}\BibitemShut {NoStop}%
\bibitem [{\citenamefont {Richaud}\ \emph {et~al.}(2019)\citenamefont {Richaud}, \citenamefont {Zenesini},\ and\ \citenamefont {Penna}}]{Richaud2019}%
  \BibitemOpen
  \bibfield  {author} {\bibinfo {author} {\bibfnamefont {A.}~\bibnamefont {Richaud}}, \bibinfo {author} {\bibfnamefont {A.}~\bibnamefont {Zenesini}},\ and\ \bibinfo {author} {\bibfnamefont {V.}~\bibnamefont {Penna}},\ }\href {https://doi.org/10.1038/s41598-019-43365-6} {\bibfield  {journal} {\bibinfo  {journal} {Scientific Reports}\ }\textbf {\bibinfo {volume} {9}},\ \bibinfo {pages} {6908} (\bibinfo {year} {2019})}\BibitemShut {NoStop}%
\bibitem [{\citenamefont {Arwas}\ \emph {et~al.}(2015)\citenamefont {Arwas}, \citenamefont {Vardi},\ and\ \citenamefont {Cohen}}]{Arwas2015}%
  \BibitemOpen
  \bibfield  {author} {\bibinfo {author} {\bibfnamefont {G.}~\bibnamefont {Arwas}}, \bibinfo {author} {\bibfnamefont {A.}~\bibnamefont {Vardi}},\ and\ \bibinfo {author} {\bibfnamefont {D.}~\bibnamefont {Cohen}},\ }\href {https://doi.org/10.1038/srep13433} {\bibfield  {journal} {\bibinfo  {journal} {Scientific Reports}\ }\textbf {\bibinfo {volume} {5}},\ \bibinfo {pages} {13433} (\bibinfo {year} {2015})}\BibitemShut {NoStop}%
\bibitem [{\citenamefont {Cataliotti}\ \emph {et~al.}(2001)\citenamefont {Cataliotti}, \citenamefont {Burger}, \citenamefont {Fort}, \citenamefont {Maddaloni}, \citenamefont {Minardi}, \citenamefont {Trombettoni}, \citenamefont {Smerzi},\ and\ \citenamefont {Inguscio}}]{Cataliotti2001}%
  \BibitemOpen
  \bibfield  {author} {\bibinfo {author} {\bibfnamefont {F.~S.}\ \bibnamefont {Cataliotti}}, \bibinfo {author} {\bibfnamefont {S.}~\bibnamefont {Burger}}, \bibinfo {author} {\bibfnamefont {C.}~\bibnamefont {Fort}}, \bibinfo {author} {\bibfnamefont {P.}~\bibnamefont {Maddaloni}}, \bibinfo {author} {\bibfnamefont {F.}~\bibnamefont {Minardi}}, \bibinfo {author} {\bibfnamefont {A.}~\bibnamefont {Trombettoni}}, \bibinfo {author} {\bibfnamefont {A.}~\bibnamefont {Smerzi}},\ and\ \bibinfo {author} {\bibfnamefont {M.}~\bibnamefont {Inguscio}},\ }\href {https://doi.org/10.1126/science.1062612} {\bibfield  {journal} {\bibinfo  {journal} {Science}\ }\textbf {\bibinfo {volume} {293}},\ \bibinfo {pages} {843} (\bibinfo {year} {2001})}\BibitemShut {NoStop}%
\bibitem [{\citenamefont {Albiez}\ \emph {et~al.}(2005)\citenamefont {Albiez}, \citenamefont {Gati}, \citenamefont {F\"olling}, \citenamefont {Hunsmann}, \citenamefont {Cristiani},\ and\ \citenamefont {Oberthaler}}]{Albiez2005}%
  \BibitemOpen
  \bibfield  {author} {\bibinfo {author} {\bibfnamefont {M.}~\bibnamefont {Albiez}}, \bibinfo {author} {\bibfnamefont {R.}~\bibnamefont {Gati}}, \bibinfo {author} {\bibfnamefont {J.}~\bibnamefont {F\"olling}}, \bibinfo {author} {\bibfnamefont {S.}~\bibnamefont {Hunsmann}}, \bibinfo {author} {\bibfnamefont {M.}~\bibnamefont {Cristiani}},\ and\ \bibinfo {author} {\bibfnamefont {M.~K.}\ \bibnamefont {Oberthaler}},\ }\href {https://doi.org/10.1103/PhysRevLett.95.010402} {\bibfield  {journal} {\bibinfo  {journal} {Phys. Rev. Lett.}\ }\textbf {\bibinfo {volume} {95}},\ \bibinfo {pages} {010402} (\bibinfo {year} {2005})}\BibitemShut {NoStop}%
\bibitem [{\citenamefont {Gati}\ and\ \citenamefont {Oberthaler}(2007)}]{Gati2007}%
  \BibitemOpen
  \bibfield  {author} {\bibinfo {author} {\bibfnamefont {R.}~\bibnamefont {Gati}}\ and\ \bibinfo {author} {\bibfnamefont {M.~K.}\ \bibnamefont {Oberthaler}},\ }\href {https://doi.org/10.1088/0953-4075/40/10/R01} {\bibfield  {journal} {\bibinfo  {journal} {Journal of Physics B: Atomic, Molecular and Optical Physics}\ }\textbf {\bibinfo {volume} {40}},\ \bibinfo {pages} {R61} (\bibinfo {year} {2007})}\BibitemShut {NoStop}%
\bibitem [{\citenamefont {Anker}\ \emph {et~al.}(2005)\citenamefont {Anker}, \citenamefont {Albiez}, \citenamefont {Gati}, \citenamefont {Hunsmann}, \citenamefont {Eiermann}, \citenamefont {Trombettoni},\ and\ \citenamefont {Oberthaler}}]{Anker2005}%
  \BibitemOpen
  \bibfield  {author} {\bibinfo {author} {\bibfnamefont {T.}~\bibnamefont {Anker}}, \bibinfo {author} {\bibfnamefont {M.}~\bibnamefont {Albiez}}, \bibinfo {author} {\bibfnamefont {R.}~\bibnamefont {Gati}}, \bibinfo {author} {\bibfnamefont {S.}~\bibnamefont {Hunsmann}}, \bibinfo {author} {\bibfnamefont {B.}~\bibnamefont {Eiermann}}, \bibinfo {author} {\bibfnamefont {A.}~\bibnamefont {Trombettoni}},\ and\ \bibinfo {author} {\bibfnamefont {M.~K.}\ \bibnamefont {Oberthaler}},\ }\href {https://doi.org/10.1103/PhysRevLett.94.020403} {\bibfield  {journal} {\bibinfo  {journal} {Physical Review Letters}\ }\textbf {\bibinfo {volume} {94}},\ \bibinfo {pages} {020403} (\bibinfo {year} {2005})}\BibitemShut {NoStop}%
\bibitem [{\citenamefont {Spagnolli}\ \emph {et~al.}(2017)\citenamefont {Spagnolli}, \citenamefont {Semeghini}, \citenamefont {Masi}, \citenamefont {Ferioli}, \citenamefont {Trenkwalder}, \citenamefont {Coop}, \citenamefont {Landini}, \citenamefont {Pezz\`e}, \citenamefont {Modugno}, \citenamefont {Inguscio}, \citenamefont {Smerzi},\ and\ \citenamefont {Fattori}}]{Spagnolli2017}%
  \BibitemOpen
  \bibfield  {author} {\bibinfo {author} {\bibfnamefont {G.}~\bibnamefont {Spagnolli}}, \bibinfo {author} {\bibfnamefont {G.}~\bibnamefont {Semeghini}}, \bibinfo {author} {\bibfnamefont {L.}~\bibnamefont {Masi}}, \bibinfo {author} {\bibfnamefont {G.}~\bibnamefont {Ferioli}}, \bibinfo {author} {\bibfnamefont {A.}~\bibnamefont {Trenkwalder}}, \bibinfo {author} {\bibfnamefont {S.}~\bibnamefont {Coop}}, \bibinfo {author} {\bibfnamefont {M.}~\bibnamefont {Landini}}, \bibinfo {author} {\bibfnamefont {L.}~\bibnamefont {Pezz\`e}}, \bibinfo {author} {\bibfnamefont {G.}~\bibnamefont {Modugno}}, \bibinfo {author} {\bibfnamefont {M.}~\bibnamefont {Inguscio}}, \bibinfo {author} {\bibfnamefont {A.}~\bibnamefont {Smerzi}},\ and\ \bibinfo {author} {\bibfnamefont {M.}~\bibnamefont {Fattori}},\ }\href {https://doi.org/10.1103/PhysRevLett.118.230403} {\bibfield  {journal} {\bibinfo  {journal} {Phys. Rev. Lett.}\ }\textbf {\bibinfo {volume} {118}},\ \bibinfo {pages} {230403} (\bibinfo {year} {2017})}\BibitemShut {NoStop}%
\bibitem [{\citenamefont {Pola}\ \emph {et~al.}(2012)\citenamefont {Pola}, \citenamefont {Stockhofe}, \citenamefont {Schmelcher},\ and\ \citenamefont {Kevrekidis}}]{Pola2012}%
  \BibitemOpen
  \bibfield  {author} {\bibinfo {author} {\bibfnamefont {M.}~\bibnamefont {Pola}}, \bibinfo {author} {\bibfnamefont {J.}~\bibnamefont {Stockhofe}}, \bibinfo {author} {\bibfnamefont {P.}~\bibnamefont {Schmelcher}},\ and\ \bibinfo {author} {\bibfnamefont {P.~G.}\ \bibnamefont {Kevrekidis}},\ }\href {https://doi.org/10.1103/PhysRevA.86.053601} {\bibfield  {journal} {\bibinfo  {journal} {Physical Review A}\ }\textbf {\bibinfo {volume} {86}},\ \bibinfo {pages} {053601} (\bibinfo {year} {2012})}\BibitemShut {NoStop}%
\bibitem [{\citenamefont {kwan Kim}\ and\ \citenamefont {Fetter}(2004)}]{Kim2004}%
  \BibitemOpen
  \bibfield  {author} {\bibinfo {author} {\bibfnamefont {J.}~\bibnamefont {kwan Kim}}\ and\ \bibinfo {author} {\bibfnamefont {A.~L.}\ \bibnamefont {Fetter}},\ }\href {https://doi.org/10.1103/PhysRevA.70.043624} {\bibfield  {journal} {\bibinfo  {journal} {Physical Review A}\ }\textbf {\bibinfo {volume} {70}},\ \bibinfo {pages} {043624} (\bibinfo {year} {2004})}\BibitemShut {NoStop}%
\bibitem [{\citenamefont {Pitaevskii}\ and\ \citenamefont {Stringari}(2016)}]{Pitaevskii2016}%
  \BibitemOpen
  \bibfield  {author} {\bibinfo {author} {\bibfnamefont {L.}~\bibnamefont {Pitaevskii}}\ and\ \bibinfo {author} {\bibfnamefont {S.}~\bibnamefont {Stringari}},\ }\href {https://doi.org/10.1093/acprof:oso/9780198758884.001.0001} {\emph {\bibinfo {title} {Bose-Einstein Condensation and Superfluidity}}}\ (\bibinfo  {publisher} {Oxford University PressOxford},\ \bibinfo {year} {2016})\BibitemShut {NoStop}%
\bibitem [{\citenamefont {Gutierrez}\ \emph {et~al.}(2021)\citenamefont {Gutierrez}, \citenamefont {de~Oliveira}, \citenamefont {Farias}, \citenamefont {Bagnato},\ and\ \citenamefont {Castilho}}]{Gutierrez2021}%
  \BibitemOpen
  \bibfield  {author} {\bibinfo {author} {\bibfnamefont {E.~M.}\ \bibnamefont {Gutierrez}}, \bibinfo {author} {\bibfnamefont {G.~A.}\ \bibnamefont {de~Oliveira}}, \bibinfo {author} {\bibfnamefont {K.~M.}\ \bibnamefont {Farias}}, \bibinfo {author} {\bibfnamefont {V.~S.}\ \bibnamefont {Bagnato}},\ and\ \bibinfo {author} {\bibfnamefont {P.~C.~M.}\ \bibnamefont {Castilho}},\ }\href {https://doi.org/10.3390/app11199099} {\bibfield  {journal} {\bibinfo  {journal} {Applied Sciences}\ }\textbf {\bibinfo {volume} {11}},\ \bibinfo {pages} {9099} (\bibinfo {year} {2021})}\BibitemShut {NoStop}%
\bibitem [{\citenamefont {Ananikian}\ and\ \citenamefont {Bergeman}(2006{\natexlab{b}})}]{Ananikian2006}%
  \BibitemOpen
  \bibfield  {author} {\bibinfo {author} {\bibfnamefont {D.}~\bibnamefont {Ananikian}}\ and\ \bibinfo {author} {\bibfnamefont {T.}~\bibnamefont {Bergeman}},\ }\href {https://doi.org/10.1103/PhysRevA.73.013604} {\bibfield  {journal} {\bibinfo  {journal} {Phys. Rev. A}\ }\textbf {\bibinfo {volume} {73}},\ \bibinfo {pages} {013604} (\bibinfo {year} {2006}{\natexlab{b}})}\BibitemShut {NoStop}%
\bibitem [{\citenamefont {Burchianti}\ \emph {et~al.}(2017)\citenamefont {Burchianti}, \citenamefont {Fort},\ and\ \citenamefont {Modugno}}]{Burchianti2017}%
  \BibitemOpen
  \bibfield  {author} {\bibinfo {author} {\bibfnamefont {A.}~\bibnamefont {Burchianti}}, \bibinfo {author} {\bibfnamefont {C.}~\bibnamefont {Fort}},\ and\ \bibinfo {author} {\bibfnamefont {M.}~\bibnamefont {Modugno}},\ }\href {https://doi.org/10.1103/PhysRevA.95.023627} {\bibfield  {journal} {\bibinfo  {journal} {Phys. Rev. A}\ }\textbf {\bibinfo {volume} {95}},\ \bibinfo {pages} {023627} (\bibinfo {year} {2017})}\BibitemShut {NoStop}%
\bibitem [{\citenamefont {Franzosi}\ and\ \citenamefont {Penna}(2003)}]{Franzosi2003}%
  \BibitemOpen
  \bibfield  {author} {\bibinfo {author} {\bibfnamefont {R.}~\bibnamefont {Franzosi}}\ and\ \bibinfo {author} {\bibfnamefont {V.}~\bibnamefont {Penna}},\ }\href {https://doi.org/10.1103/PhysRevE.67.046227} {\bibfield  {journal} {\bibinfo  {journal} {Physical Review E}\ }\textbf {\bibinfo {volume} {67}},\ \bibinfo {pages} {046227} (\bibinfo {year} {2003})}\BibitemShut {NoStop}%
\bibitem [{\citenamefont {Mossmann}\ and\ \citenamefont {Jung}(2006)}]{Mossmann2006}%
  \BibitemOpen
  \bibfield  {author} {\bibinfo {author} {\bibfnamefont {S.}~\bibnamefont {Mossmann}}\ and\ \bibinfo {author} {\bibfnamefont {C.}~\bibnamefont {Jung}},\ }\href {https://doi.org/10.1103/PhysRevA.74.033601} {\bibfield  {journal} {\bibinfo  {journal} {Phys. Rev. A}\ }\textbf {\bibinfo {volume} {74}},\ \bibinfo {pages} {033601} (\bibinfo {year} {2006})}\BibitemShut {NoStop}%
\bibitem [{\citenamefont {Hernández-Rajkov}\ \emph {et~al.}(2024)\citenamefont {Hernández-Rajkov}, \citenamefont {Grani}, \citenamefont {Scazza}, \citenamefont {Pace}, \citenamefont {Kwon}, \citenamefont {Inguscio}, \citenamefont {Xhani}, \citenamefont {Fort}, \citenamefont {Modugno}, \citenamefont {Marino},\ and\ \citenamefont {Roati}}]{Hernández-Rajkov2024}%
  \BibitemOpen
  \bibfield  {author} {\bibinfo {author} {\bibfnamefont {D.}~\bibnamefont {Hernández-Rajkov}}, \bibinfo {author} {\bibfnamefont {N.}~\bibnamefont {Grani}}, \bibinfo {author} {\bibfnamefont {F.}~\bibnamefont {Scazza}}, \bibinfo {author} {\bibfnamefont {G.~D.}\ \bibnamefont {Pace}}, \bibinfo {author} {\bibfnamefont {W.~J.}\ \bibnamefont {Kwon}}, \bibinfo {author} {\bibfnamefont {M.}~\bibnamefont {Inguscio}}, \bibinfo {author} {\bibfnamefont {K.}~\bibnamefont {Xhani}}, \bibinfo {author} {\bibfnamefont {C.}~\bibnamefont {Fort}}, \bibinfo {author} {\bibfnamefont {M.}~\bibnamefont {Modugno}}, \bibinfo {author} {\bibfnamefont {F.}~\bibnamefont {Marino}},\ and\ \bibinfo {author} {\bibfnamefont {G.}~\bibnamefont {Roati}},\ }\href {https://doi.org/10.1038/s41567-024-02466-4} {\bibfield  {journal} {\bibinfo  {journal} {Nature Physics}\ }\textbf {\bibinfo {volume} {20}},\ \bibinfo {pages} {939} (\bibinfo {year} {2024})}\BibitemShut {NoStop}%
\bibitem [{\citenamefont {Caldara}\ \emph {et~al.}(2024)\citenamefont {Caldara}, \citenamefont {Richaud}, \citenamefont {Capone},\ and\ \citenamefont {Massignan}}]{Caldara2024}%
  \BibitemOpen
  \bibfield  {author} {\bibinfo {author} {\bibfnamefont {M.}~\bibnamefont {Caldara}}, \bibinfo {author} {\bibfnamefont {A.}~\bibnamefont {Richaud}}, \bibinfo {author} {\bibfnamefont {M.}~\bibnamefont {Capone}},\ and\ \bibinfo {author} {\bibfnamefont {P.}~\bibnamefont {Massignan}},\ }\href@noop {} {\  (\bibinfo {year} {2024})},\ \Eprint {https://arxiv.org/abs/2403.11987} {arXiv:2403.11987 [cond-mat.quant-gas]} \BibitemShut {NoStop}%
\bibitem [{\citenamefont {Prates}\ \emph {et~al.}(2022)\citenamefont {Prates}, \citenamefont {Zezyulin},\ and\ \citenamefont {Konotop}}]{Prates2022}%
  \BibitemOpen
  \bibfield  {author} {\bibinfo {author} {\bibfnamefont {H.~C.}\ \bibnamefont {Prates}}, \bibinfo {author} {\bibfnamefont {D.~A.}\ \bibnamefont {Zezyulin}},\ and\ \bibinfo {author} {\bibfnamefont {V.~V.}\ \bibnamefont {Konotop}},\ }\href {https://doi.org/10.1103/PhysRevResearch.4.033219} {\bibfield  {journal} {\bibinfo  {journal} {Phys. Rev. Res.}\ }\textbf {\bibinfo {volume} {4}},\ \bibinfo {pages} {033219} (\bibinfo {year} {2022})}\BibitemShut {NoStop}%
\bibitem [{\citenamefont {Pigneur}\ and\ \citenamefont {Schmiedmayer}(2018)}]{Pigneur2018}%
  \BibitemOpen
  \bibfield  {author} {\bibinfo {author} {\bibfnamefont {M.}~\bibnamefont {Pigneur}}\ and\ \bibinfo {author} {\bibfnamefont {J.}~\bibnamefont {Schmiedmayer}},\ }\href {https://doi.org/10.1103/PhysRevA.98.063632} {\bibfield  {journal} {\bibinfo  {journal} {Physical Review A}\ }\textbf {\bibinfo {volume} {98}},\ \bibinfo {pages} {063632} (\bibinfo {year} {2018})}\BibitemShut {NoStop}%
\end{thebibliography}
\end{document}